\documentclass[onecolumn,noshowpacs,superscriptaddress,nobibnotes,nofootinbib,12pt,showkeys,preprintnumbers]{revtex4-1}
\UseRawInputEncoding
\usepackage{float}
\usepackage{amsmath,amssymb}
\usepackage{bm}
\usepackage{slashed}
\usepackage{epsfig}
\usepackage{graphicx}
\usepackage{hyperref}
\usepackage{xcolor}
\hypersetup{
    colorlinks=true,
    linkcolor=blue,
    filecolor=magenta,      
    urlcolor=blue,
    citecolor=blue
}

\begin{document}
\title{Factorization and transverse phase-space parton distributions}
\author{Bin Wu}
\email{b.wu@cern.ch}
\affiliation{Theoretical Physics Department, CERN\\ CH-1211 Geneva 23, Switzerland}
\email{b.wu@cern.ch}
\affiliation{Instituto Galego de F\'isica de Altas Enerx\'ias IGFAE, Universidade de Santiago de Compostela,
E-15782 Galicia-Spain}
\affiliation{LIP, Av. Prof. Gama Pinto, 2, P-1649-003 Lisboa, Portugal}

\begin{abstract}
We first revisit impact-parameter dependent collisions of ultra-relativistic particles in quantum field theory. Two conditions sufficient for defining an impact-parameter dependent cross section are given, which could be violated in proton-proton collisions. By imposing these conditions, a general formula for the impact-parameter dependent cross section is derived.
Then, using soft-collinear effective theory, we derive a factorization formula for the impact-parameter dependent cross section for inclusive hard processes with only colorless final-state products in hadron and nuclear collisions. It entails defining thickness beam functions, which are Fourier transforms of transverse phase-space parton distribution functions. By modelling non-perturbative modes in thickness beam functions of large nuclei in heavy-ion collisions, the factorization formula confirms the cross section in the Glauber model for hard processes. Besides, the factorization formula is verified up to one loop in perturbative QCD for the inclusive Drell-Yan process in quark-antiquark collisions at a finite impact parameter.
\end{abstract}
\preprint{CERN-TH-2021-019}

\maketitle

\section{Introduction}
\label{sec:intro}
In hadron and nuclear collisions, cross sections for hard processes are always a blend of short-distance and long-distance physics. QCD factorization allows one to systematically study incoherence between effects at various distance scales~\cite{Collins:1989gx}. The predictive power of perturbative QCD for hadron collider physics relies on validity of factorization theorems and universality of quantities encoding long-distance physics. Factorization also provides a systematic approach to resum large logarithms of ratios between different scales to all orders in perturbation theory~\cite{Sterman:1995fz}. 

Factorization for the Drell-Yan cross section in hadron collisions has been extensively studied in the literature. Its validity has been proved within the context of perturbative QCD~\cite{Bodwin:1984hc,Collins:1984kg,Collins:1985ue,Collins:1988ig}. It has been alternatively studied using the soft-collinear effective theory (SCET)~\cite{Bauer:2000yr,Bauer:2001ct,Bauer:2001yt,Bauer:2002nz,Beneke:2002ph} in Refs. \cite{Bauer:2002nz, Becher:2010tm}. Such an effective field theory approach facilitates exploring factorization and resummation in other similar processes, such as inclusive Higgs boson production~\cite{Mantry:2009qz}.

Long-distance physics in all the processes mentioned above manifests itself either in parton distribution functions (PDFs)~\cite{Curci:1980uw, Collins:1981uw}, transverse-momentum-dependent (TMD) PDFs~\cite{Collins:1981uw} or the beam functions~\cite{Stewart:2009yx} in SCET. They are all defined as matrix elements of gauge invariant operators sandwiched between some momentum eigenstate of colliding hadrons. And they do not possess any information on the spatial distribution of quarks and gluons inside the hadrons, needed for a holistic snapshot of quantum phases-space parton distributions~\cite{Belitsky:2003nz}.

The spatial distribution of partons in a hadron, say, a proton, could be revealed by studying impact-parameter dependent collisions, as have been extensively studied in nucleus-nucleus (AA) collisions. The impact parameter in AA/heavy-ion collisions can be determined via centrality measurements by using the Glauber model~\cite{Miller:2007ri}. In this model, the cross section for the Drell-Yan production of vector bosons reduces to that in binary nucleon-nucleon collisions, which is consistent with recent measurements at the LHC within experimental uncertainties~\cite{Aad:2010aa, Chatrchyan:2011ua, Aad:2012ew,Chatrchyan:2012nt, Chatrchyan:2012vq}. In order to unambiguously define the impact parameter of proton-proton (pp) collisions, one needs first to clarify conceptual difference between small colliding systems like pp collisions and large colliding systems like AA collisions.

The discovery of collectivity in pp collisions~\cite{Khachatryan:2010gv,Aad:2015gqa,Khachatryan:2015lva, Khachatryan:2016txc,Aaboud:2017blb,Sirunyan:2017uyl}, however, blurs the boundary between large and small colliding systems~\cite{Loizides:2016tew}. Concepts based on (classical) collision geometry and the impact parameter in heavy-ion collisions have been frequently employed to interpret collectivity in pp collisions in many theoretical discussions without scrutiny (see Ref. \cite{Nagle:2018nvi} for a recent review). Nowadays, redefining the boundary of physical concepts respectively applicable to pp, pA and AA collisions is one of the main focuses in high-energy nuclear physics~\cite{Loizides:2016tew, Nagle:2018nvi}, which demands a unified theoretical approach in QCD to treat all these collisions on the same footing. 

\begin{figure}
\includegraphics[width=0.7\textwidth]{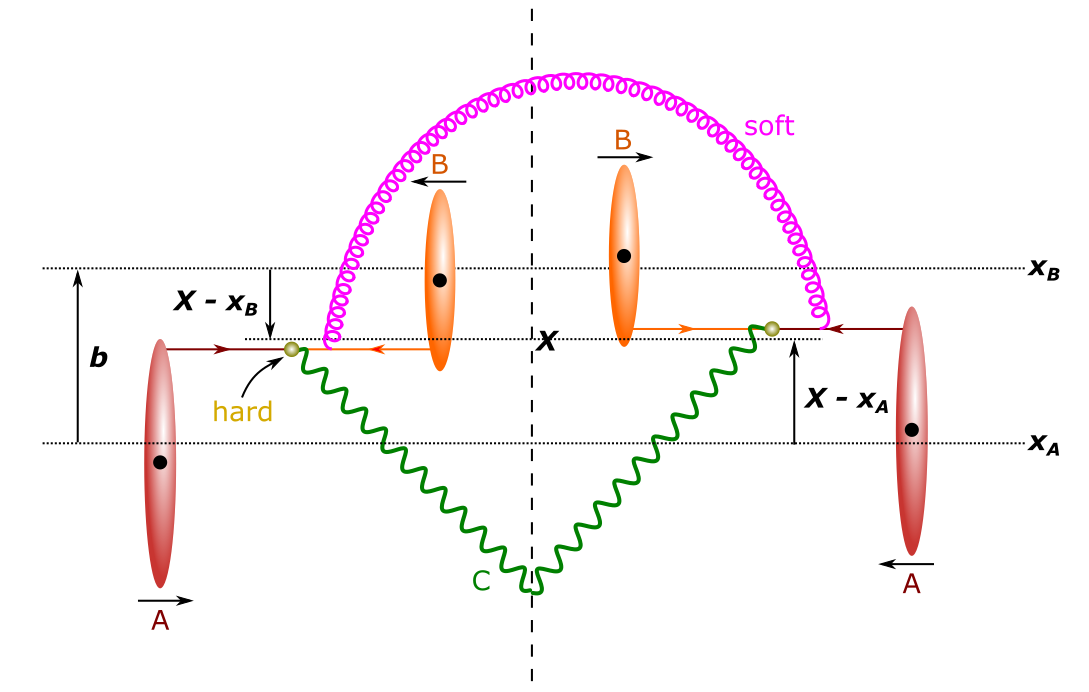}
\caption{The quantum picture of impact-parameter dependent collisions. This figure combines the amplitude and the conjugate amplitude for the hard process in Eq. (\ref{eq:process}) into a cut diagram. The transverse momenta of particles $A$ and $B$ are known to be around some design values with an uncertainty $\Delta p_T$. Accordingly, their transverse locations $\mathbf{x}_A$ and $\mathbf{x}_B=\mathbf{x}_A+\mathbf{b}$ can only be determined with an accuracy $\Delta x_T$ limited by the uncertainty principle. This accounts for the displacement of their transverse positions across the cut.  The product $C$ with momentum $p_C^\mu$ is created at some hard scattering vertex around $\mathbf{X}$, initiated by two partons respectively collinear to $A$ and $B$. Different collinear partons can communicate with one another via exchange of soft gluons. The misalignment of the hard vertex in the transverse plane across the cut $\mathbf{x}$, that is, its transverse coherence length, is of order $1/p_{C,T}$.
\label{fig:sigmb}}
\end{figure} 

The sole purpose  of this paper is to lay the groundwork for a unified description of hard processes in impact-parameter dependent pp, pA and AA collisions based on QCD factorization. As depicted in Fig. \ref{fig:sigmb}, we restrict ourselves to a generic inclusive hard process in hadron and nuclear collisions at an impact parameter $b$:
\begin{align}\label{eq:process}
    A+B\to C + \text{anything else}
\end{align}
with $A$ and $B$ either hadrons or nuclei, and $C$ some colorless final-state object, such as electroweak gauge bosons and Higgs bosons. Below, we motivate and outline how we proceed to derive the factorized cross section for such a hard process.

In Sec. \ref{sec:cross_section}, we revisit the fuzzy quantum picture of impact-parameter dependent collisions. The motivation for this section is two-fold: First,
in a typical pp collision at the LHC, the proton size $R$, the impact parameter $b$ and even the proton transverse spatial dispersion $\Delta x_T$\footnote{
At the LHC, the crossing angle at the interaction point $\theta_c$ can be measured with an accuracy $\Delta \theta_c < 10~\mu$rad~\cite{Evans:2008zzb}. For a $E = 7$ TeV proton, the uncertainty principle dictates $\Delta x_T\geq \frac{1}{2 E\Delta\theta_c}>1.4$ fm.}
could be comparable with one another. In this case, a quantum description of the collision becomes more appropriate. Second, in heavy-ion collisions, the Glauber model \cite{Miller:2007ri} has  been broadly employed not only in theoretical calculations but also in experimental studies of collision geometry from centrality measurements. Behind the Glauber model underlies a classical picture of the impact parameter, which can even trace back to  Rutherford's seminal discovery in Ref. \cite{Rutherford:1911zz} that helped usher in the quantum age. Giving this model a justification in QCD entails a quantum picture instead. However, the formula for the impact-parameter dependent cross section has not been derived in quantum scattering theory~\cite{Taylor:1972,Peskin:1995ev}. This section is aimed to fill this gap.

We consider a collision of two ultra-relativistic particles $A$ and $B$ at the impact parameter $b$. In collider physics, all we know is that their momenta are determined to be around some design values $P_i^\mu$ with $i=A$ and $B$. From this information, we do know that the states of the particles can be described by some wave packets in momentum space, with their momentum width constrained by experimental uncertainties. Accordingly, the particles are localized in position space, described by some well-defined spatial wave packets. We aim to define the impact-parameter dependent cross section for the collision, which is intrinsic to the colliding particles and should be independent of the beam particles' wave packets.

We find that such an impact-parameter dependent cross section can be defined, as given in Eq. (\ref{eq:dsigmadbdo}), if the following two conditions are fulfilled. Condition i) is the high-energy limit: $|P_{iz}|\gg P_{iT}, \Delta p_T, \Delta p_z$ with $\Delta p_T$ and $\Delta p_z$ respectively the particle's transverse and longitudinal momentum dispersions. It is needed to identify the longitudinal momenta of the wave packets in the amplitude and the conjugate amplitude. Condition ii) is to require that the fuzziness in particles' transverse positions should be smaller than $b$: $b\gg \Delta x_T$. It is needed to identify the transverse momenta of the wave packets in the amplitude and the conjugate amplitude. Therefore, they are both necessary for integrating out the wave packets in order to define the cross section and for unambiguously defining the impact parameter. Otherwise, the probability for producing any final states in the collision explicitly depends on the wave packets, which are not guaranteed to be the same in different experiments. And, one ends up measuring a quantity which differs across experiments.

With the general formula for the impact-parameter dependent cross section derived in Sec. \ref{sec:cross_section}, one can calculate it for the hard process in Eq. (\ref{eq:process}) using perturbative QCD. In general, one needs to deal with the long-distance behavior in perturbative series. We appeal to QCD factorization to factor out long-distance physics in Sec. \ref{sec:nuclear}. 

In Sec. \ref{sec:factorization}, using SCET we first give a detailed derivation of the factorized cross section for the process in Eq. (\ref{eq:process}). The main features of this factorization formula can be understood based on the following heuristic argument:

Imagine that besides the hard scale $Q$ we measure the transverse momentum $p_{C,T}\gg \Lambda_{QCD}$ of the product $C$. As illustrated in Fig. \ref{fig:sigmb}, $C$ is produced at some hard scattering vertex around $\mathbf{X}$ with a transverse coherent length $|\mathbf{x}|\sim 1/p_{C,T}\ll 1$ fm within a time scale $t_h\sim 1/Q$. It effectively picks out two partons located around $\mathbf{X}$ respectively from $A$ and $B$ with a transverse spatial accuracy of the order of $|\mathbf{x}|$. The distributions of these partons in the colliding particles are described by a new type of PDFs, which are referred to as {\it thickness beam functions} in this paper.

The thickness beam functions are expected to be universal as a consequence of the hard-collinear and soft-collinear factorization. Generic hard processes all involve radiation collinear to the two beam directions $n_A^\mu$ and $n_B^\mu$. The coherent time of a $n_i$-collinear parton with momentum $p_{n_i}^{\mu}$ is given by 
\begin{align}
t_{n_i}=\frac{1}{n_i\cdot p_{n_i}}
\sim \frac{1}{\lambda^2 Q}\gg t_h\sim\frac{1}{Q}
\end{align}
where the expansion parameter $\lambda$, as a ratio between a soft momentum scale determined by some specific observable to be measured and the hard scale $Q$, is assumed to be much smaller than unity. That is, the beam collinear radiation takes place too early to interfere with the hard scattering. 
Different beam collinear partons may communicate, via exchange of soft gluons, with each other or with final-state colored objects since soft gluons have a coherent time $t_s\gg t_h$:
\begin{align}
    t_s\sim \frac{1}{\lambda Q},\qquad \text{or}\qquad\frac{1}{\lambda^2 Q}.
\end{align}
As long as soft gluons cannot resolve the substructure of collinear splittings, there is factorization between soft and beam collinear partons. And the thickness beam functions should be universal to all the hard processes in which these two types of factorization hold true.

Because their formation time is much longer than $t_h$, the hard scattering occurs too rapidly to radiate soft gluons. That is, soft gluons are detached from the hard scattering vertex. As a consequence, for the process in Eq. (\ref{eq:process}), soft radiation only couples to the total color charges of collinear partons from $A$ or $B$ and organizes itself into soft Wilson lines. They act on the vacuum state to define another module of the factorization formula: the soft function. Since it is the vacuum state that is involved in its definition, the soft function is independent of $\mathbf{X}$ and does not carry any information on the impact parameter.

Based on the above argument one can expect that the factorization formula, as given in Eq. (\ref{eq:factorization}), schematically takes the following form:
\begin{align}\label{eq:factorizationIntro}
   \frac{d\sigma_{AB}}{d^2\mathbf{b} dy_C d^2\mathbf{p}_C} 
\sim & \sum\limits_{j,k}\int d^2\mathbf{X}\mathcal{T}_{j/A}(\mathbf{X}-\mathbf{x}_A) \otimes\mathcal{T}_{k/B}(\mathbf{X} -\mathbf{x}_B)\otimes H_{j,k\to C}\otimes \mathcal{S}
\end{align}
with the impact parameter $\mathbf{b}=\mathbf{x}_B-\mathbf{x}_A$, $\mathbf{X}$ denoting the transverse position of the hard scattering and $\mathbf{x}_i$ being the (average) transverse location of particle $i$, as shown in Fig. \ref{fig:sigmb}. Here, the hard function $H_{j,k\to C}$ can be calculated from the partonic process $j, k\to C + \text{anything else}$ and $\mathcal{S}$ is the soft function. They are the same as for conventional cross sections in pp collisions~\cite{Bauer:2002nz, Becher:2010tm,Mantry:2009qz}.  The thickness beam functions $\mathcal{T}_{j/i}$ are related to the quantum transverse phase-phase PDFs via the Fourier transform. That is, they encode the information on the distribution of parton $j$ carrying a momentum fraction $z$ in both transverse momentum and position spaces inside particle $i$. A detailed discussion about the properties of $\mathcal{T}_{j/i}$ is presented in Sec. \ref{sec:T}.

In Sec. \ref{sec:glauber}, we study the connection between the factorization formula in Eq. (\ref{eq:factorizationIntro}) and the cross section in the Glauber model in heavy-ion collisions. We find that $\mathcal{T}_{j/i}$ reduces to a product of the thickness function in the Glauber model and
the corresponding beam function, as the Fourier transform of the TMD PDF, when the incoming nucleus is treated as an assembly of uncorrelated nucleons. This gives the success of the Glauber model~\cite{dEnterria:2020dwq} a QCD justification for such hard processes. On the other hand, the factorization formula allows one to explore refined details about the parton distributions in heavy nuclei. It can be used to optimize the potential of the LHC and HL-LHC, with increased accuracy, in systematically studying cold nuclear effects~\cite{Vogt:2000hp,Armesto:2006ph, Eskola:2009uj}, which are absent in the aforementioned modelling. Nuclear modifications of PDFs had been first revealed in deep inelastic scattering by the European Muon Collaboration~\cite{Aubert:1983xm} and have been favored by recent measurements in PbPb collisions at the LHC~\cite{Sirunyan:2019dox}. The factorization formula is important for investigating the LHC/HL-LHC's potential to pin down such effects since the hard and soft functions can be calculated at high accuracy in perturbative QCD.

Using the factorization formula in Eq. (\ref{eq:factorizationIntro}), one can potentially measure the transverse phase-space parton distributions in protons through inclusive hard processes in impact-parameter dependent pp collisions, which have not been experimentally explored. On the other hand, given the fact that the aforementioned two conditions needed to define the impact-parameter dependent collisions are not always fulfilled, caution is needed when the impact-parameter dependent cross section is studied in pp collisions. A brief discussion on this issue is given in Sec. \ref{sec:pp}. 

At this point, we have two ways to calculate the impact-parameter dependent cross section by either using the general formula in Sec. \ref{sec:cross_section} or the factorization formula in Sec. \ref{sec:nuclear}. This provides a way to verify factorization order by order in perturbation theory: first, calculate the cross section in perturbative QCD using the general formula; then, expand the perturbative QCD results at small $\lambda$ or, equivalently, large $Q$; and finally compare the leading-order results in $Q$ to those given by  the factorization formula. 

In Sec. \ref{sec:qqb}, a verification of the factorization formula is carried out for the inclusive Drell-Yan process $q\bar{q}\to \gamma^*$. This process involves two scales: the photon virtuality $Q$ and the impact parameter $b$, which set the expansion parameter $\lambda=1/(bQ)$. We only focus on the physically interesting case with $\lambda \ll 1$, that is, the impact parameter being much larger than $1/Q$. At leading order in $\lambda$, the factorization formula is confirmed up to one loop in perturbative QCD. We also calculate the one-loop spatial quark distribution in a fast moving quark, which carries information complementary to the corresponding TMD PDF. The probability to find a quark carrying a momentum fraction $z$ at a transverse distance $r$ from the original transverse location of the incoming quark is found to be inversely proportional to  $r^2$.

Detailed derivations and calculations backing  up the above summary can be found in the ensuing sections. And the interested reader is invited to vet their details.

\section{The impact-parameter dependent cross section}
\label{sec:cross_section}

In this section we revisit the concepts of the impact parameter and the impact-parameter dependent cross section in quantum field theory (QFT). These two concepts are well-defined in classical physics, as exemplified by the original derivation of the Rutherford scattering formula~\cite{Rutherford:1911zz}: the deflection angle of a charged particle scattering off a Coulomb potential is given by a unique function of the impact parameter. In contrast, in the textbook derivation of the conventional cross section (see, e.g., Chapter 4 of \cite{Peskin:1995ev}), there is no unique relation between the deflection angle and the impact parameter. Yet, Rutherford's formula can be easily reproduced. Below, we redo the derivation of the formula for the cross section in QFT in order to restore its impact-parameter dependence.

Consider a collision of two ultra-relativistic particles $A$ and $B$ respectively from two counter-moving beams. Before the collision, the momentum of particle $i$ with $i=A$ or $B$ is accelerated to be around some design value $\vec{P}_i$ with an uncertainty $\Delta \vec{p}_i$. So, all we know is that its wave packet peaks about $\vec{P}_i$ with a momentum width equal to or smaller than $\Delta \vec{p}_i$, which can be generically written in the form\footnote{
We only focus on unpolarized collisions and ignore the wave packet's spin dependence here. 
}
\begin{align}\label{eq:wavepacket}
    |\phi_i\rangle &= \int\frac{d^3\vec{p}}{(2\pi)^3}\frac{e^{-i\mathbf{p}\cdot \mathbf{x}_i}}{\sqrt{2 E_p}}\phi_i(\vec{p})|\vec{p}\rangle,
\end{align}
where $\phi_i$ is normalized, that is,
\begin{align}
    1=\langle\phi_i|\phi_i\rangle=\int\frac{d^3\vec{p}}{(2\pi)^3}|\phi_i(\vec{p})|^2,
\end{align}
the phase factor in front of $\phi_i(\vec{p})$ accounts for the spatial translation in the transverse plane and $\mathbf{x}_i$ is the transverse position vector of particle $i$, to be determined later. Here and below, we denote two-dimensional vectors in the transverse plane by bold letters and three-dimensional vectors by letters with an arrow overhead. 

The impact parameter of the collision is defined as
\begin{align}\label{eq:b}
    {\mathbf b}\equiv {\mathbf x}_B - {\mathbf x}_A,
\end{align}
once $\mathbf{x}_A$ and $\mathbf{x}_B$ are given. However, the particles are, at best, known to locate somewhere with an uncertainty limited by the uncertainty principle. In order to quantitatively define $\mathbf{x}_i$, one needs the particle's spatial wave packet, which is given by
\begin{align}\label{eq:phis}
    \tilde\phi_i(\mathbf{x}-\mathbf{x}_i, z) =\langle\vec{x}|\phi_i \rangle\equiv  \int\frac{d^3\vec{p}}{(2\pi)^3}e^{i\vec{p}\cdot\vec{x}-i\mathbf{p}\cdot \mathbf{x}_i}\phi_i(\vec{p})
\end{align}
with the position eigenstate~\cite{Newton:1949cq}
\begin{align}
    |\vec{x}\rangle &\equiv \int\frac{d^3\vec{p}}{(2\pi)^3}\frac{e^{-i\vec{p}\cdot \vec{x}}}{\sqrt{2 E_p}}|\vec{p}\rangle.
\end{align}
As long as $\vec{x}$ is not measured with a resolution better than the particle's de Broglie wavelength, $|\tilde\phi_i|^2$ admits interpretation as the probability density to find particle $i$ at $\vec{x}$. Given $\tilde{\phi}_i$, $\mathbf{x}_i$ can be chosen to be the average transverse position vector of particle $i$. Obviously, there are alternative choices for $\mathbf{x}_i$, such as the center of mass or the transverse peak location of $|\tilde\phi_i|^2$. Below, we show that such an ambiguity in defining $b\equiv |\mathbf{b}|$ is negligible when one is allowed to define an impact-parameter dependent cross section. 

In principle, one can predict the probability for producing any final state $|\{p_f\}\rangle$ in the collision at the impact parameter $\mathbf{b}$ according to
\begin{align}\label{eq:Pbdef}
    P_{\mathbf{b}}(\phi_A, \phi_B\to \{ p_f\})&=\langle \phi_A\phi_B|\hat{S}^\dagger|\{ p_f\}\rangle\langle \{ p_f\}|\hat{S}|\phi_A \phi_B\rangle.
\end{align}
Since we are only interested in the cross section, we can replace the $S$-matrix element by
\begin{align}
    \langle \{ p_f\}|\hat{S}|p_A, p_B\rangle \to (2\pi)^4 \delta^{(4)}(p_A+p_B-\sum p_f)i M(p_A, p_B\to \{ p_f\}).
\end{align}
Then, plugging the wave packets in Eq. (\ref{eq:wavepacket}) into Eq. (\ref{eq:Pbdef}) and using one of the delta functions from the above replacement to integrate out $\bar{p}_{Az},\bar{p}_{Bz}$ and $\mathbf{\bar{p}}_B$ yields
\begin{align}\label{eq:dPdb_full}
    P_{\mathbf{b}}(\phi_A, \phi_B\to \{ p_f\})=&\int\limits_{\mathbf{p}_A, \mathbf{\bar{p}}_A, \mathbf{p}_B, p_{Az}, p_{Bz}} e^{i\mathbf{b}\cdot (\mathbf{p}_A - \mathbf{\bar{p}}_A)} \notag\\
    &\times\phi_A(\vec{p}_A)\phi^*_A(\vec{\bar{p}}_A)\phi_B(\vec{p}_B)\phi^*_B(\vec{\bar{p}}_B)\sigma(p_A, p_B\to \{ p_f\} \leftarrow\bar{p}_A, \bar{p}_B),
\end{align}
where the measure $\int\prod\limits_{j=1}^n\frac{dp_j}{2\pi}$ is denoted by $\int\limits_{p_1,\cdots,p_n}$ for brevity, the off-diagonal cross section is defined as
\begin{align}\label{eq:offsigma}
    \sigma(p_A, p_B\to \{ p_f\} \leftarrow\bar{p}_A, \bar{p}_B)\equiv&\frac{(2\pi)^4\delta^{(4)}(p_A + p_B - \sum p_f)}{\sqrt{2E_{p_A}2E_{\bar{p}_A}2E_{p_B}2E_{\bar{p}_B}}|\bar{v}_{Az} - \bar{v}_{Bz}|}\notag\\
    &\times M(p_A, p_B \to \{p_f\}) M^*(\bar{p}_A, \bar{p}_B \to \{p_f\}),
\end{align}
$\mathbf{\bar{p}}_B = \mathbf{{p}}_A + \mathbf{{p}}_B - \mathbf{{\bar{p}}}_A$, and the longitudinal momenta $\bar{p}_{Az}$ and $\bar{p}_{Bz}$ are solutions to
\begin{align}\label{eq:pz}
    E_{p_A} + E_{p_B} = E_{\bar{p}_A} + E_{\bar{p}_B},\qquad \bar{p}_{Az} + \bar{p}_{Bz} = p_{Az} + p_{Bz}.
\end{align}

The impact-parameter dependent probability $P_{\mathbf{b}}$ generally depends on the wave packets. However, the beam particles' wave packets are not measured in collider physics. Moreover, they are not guaranteed to be the same in different experiments. The cross section, on the other hand, is intrinsic to the colliding particles and, therefore, should be independent of the wave packets in order to allow comparison across experiments. Below, we show that the following two conditions
\begin{align}\label{eq:cond}
    \text{i) }|P_{iz}|\gg|\mathbf{P}_i|, \Delta p_T, \Delta p_{z}; \qquad \text{ii)} |\mathbf{b}|\gg \Delta x_T
\end{align}
are sufficient for defining the impact-parameter dependent cross section from $P_{\mathbf{b}}$. Here, $\Delta x_T$, $\Delta p_T$ and $\Delta p_z$ are respectively the transverse spatial, transverse and longitudinal momentum dispersions of the colliding particles.

Condition i) is the high-energy limit in which both particles are moving predominantly along the beam ($\pm$z) directions. In this limit, the solutions to Eq. (\ref{eq:pz}) are given by
\begin{align}\label{eq:dpz}
    \bar{p}_{Az} - p_{Az} =  p_{Bz}-\bar{p}_{Bz}\approx\frac{|\mathbf{p}_{A}|^2-|\mathbf{\bar{p}}_{A}|^2}{4 P_{Az}} - \frac{|\mathbf{p}_{B}|^2-|\mathbf{\bar{p}}_{B}|^2}{4 P_{Bz}}
\end{align}
for $P_{Az}>0$ and $P_{Bz}<0$. Since these terms are small\footnote{
For example, estimated from crossing angles at the LHC~\cite{Evans:2008zzb}, one has $|\bar{p}_{iz}-p_{iz}|/|P_{iz}|<10^{-10}$.
}, we will drop them and take
\begin{align}\label{eq:cond1}
\bar{p}_{Az} = p_{Az},\qquad \bar{p}_{Bz} = p_{Bz}.
\end{align}
Following \cite{Wigner:1932eb}, we define transverse Wigner functions
\begin{align}\label{eq:WT}
    W_i(\mathbf{X}, \mathcal{P})\equiv \int\frac{d^2\mathbf{\chi}}{(2\pi)^2}\int dz e^{i \mathcal{P}\cdot \mathbf{\chi}}&\tilde\phi_i\left(\mathbf{X}-\frac{\mathbf{\chi}}{2}, z\right)\tilde\phi_i^*\left(\mathbf{X}+\frac{\mathbf{\chi}}{2}, z\right).
\end{align}
In terms of $W_i$, $P_{\mathbf{b}}$ can be expressed as
\begin{align}\label{eq:dPdb}
    P_{\mathbf{b}}(\phi_A, \phi_B\to \{ p_f\})  =&
    \int d^2\mathbf{X}_A d^2\mathbf{\mathcal{P}}_A d^2\mathbf{X}_B d^2\mathbf{\mathcal{P}}_B\int \frac{d^2\mathbf{q}}{(2\pi)^2}e^{i\mathbf{q}\cdot\left(\mathbf{b} + \mathbf{X}_B - \mathbf{X}_A\right)}\notag\\
    &\times W_A(\mathbf{X}_A, \mathbf{\mathcal{P}}_A)W_B(\mathbf{X}_B, \mathbf{\mathcal{P}}_B)\sigma(p_A, p_B\to \{ p_f\} \leftarrow\bar{p}_A, \bar{p}_B)
\end{align}
with
\begin{align}
    \mathbf{p}_A&=\mathcal{P}_A+\frac{\mathbf{q}}{2},\qquad
      \mathbf{p}_B=\mathcal{{P}}_B - \frac{\mathbf{q}}{2},\notag\\
     \bar{\mathbf{p}}_A&=\mathcal{{P}}_A-\frac{\mathbf{q}}{2}
     ,\qquad
      \bar{\mathbf{p}}_B=\mathcal{{P}}_B+\frac{\mathbf{q}}{2}.
\end{align}

Given condition i), one is allowed to expand the off-diagonal cross section in Eq. (\ref{eq:offsigma}) at large $P_{iz}$. For collider phenomenology, one only needs to keep leading-order terms in $\Delta/P_{iz}$ with $\Delta = \mathcal{P}_i, \Delta p_T$ or $\Delta p_z$, which is equivalent to
replacing $\mathbf{\mathcal{P}}_i$ by $\mathbf{0}$ and $p_{iz}$ by their design values $P_{iz}$ in the off-diagonal cross section\footnote{
Another justification of this approximation is that modern detectors typically have limited resolution which can not resolve such a small variation of $P_i$~\cite{Peskin:1995ev}.
}. $|\mathbf{q}|\sim 1/b$ could also be much smaller than $|P_{iz}|$. However, as it will become clear in the following sections, we do not expand the amplitude squared in the off-diagonal cross section about $|\mathbf{q}|/P_{iz}=0$ here. We only do so when it is needed in detailed calculations, as exemplified in Sec. \ref{sec:qqb}. As a result, we have
\begin{align}\label{eq:offsigmaLO}
    \sigma(p_A, p_B\to \{ p_f\} \leftarrow\bar{p}_A, \bar{p}_B)=&\frac{1}{2s}M(p_A, p_B \to \{p_f\}) M^*(\bar{p}_A, \bar{p}_B \to \{p_f\})\notag\\
    &\times (2\pi)^4\delta^{(4)}(p_A + p_B - \sum p_f),
\end{align}
where the Mandelstam variable
\begin{align}
    s=\bar{n}_A\cdot P_A \bar{n}_B\cdot P_B,
\end{align}
the incoming momenta are given by
\begin{align}\label{eq:psincoming}
    &p_A^\mu = \bar{n}_A\cdot P_A\frac{n_A^\mu}{2}+ \frac{q_T^\mu}{2}- \frac{q_T^2}{4 \bar{n}_A\cdot P_A} \frac{\bar{n}_A^\mu}{2},\qquad p_B^\mu = \bar{n}_B\cdot P_B\frac{n_B^\mu}{2}- \frac{q_T^\mu}{2}- \frac{q_T^2}{4 \bar{n}_B\cdot P_B} \frac{\bar{n}_B^\mu}{2},\notag\\
    &\bar{p}_A^\mu = \bar{n}_A\cdot P_A\frac{n_A^\mu}{2}- \frac{q_T^\mu}{2}- \frac{q_T^2}{4 \bar{n}_A\cdot P_A} \frac{\bar{n}_A^\mu}{2},\qquad \bar{p}_B^\mu =\bar{n}_B\cdot P_B\frac{n_B^\mu}{2}+ \frac{q_T^\mu}{2}- \frac{q_T^2}{4 \bar{n}_B\cdot P_B} \frac{\bar{n}_B^\mu}{2},
\end{align}
with masses being neglected,
and for any four-vector $V^\mu$ we define
\begin{align}
    V_T^\mu={{g_T}^{\mu}}_{\nu} V^\nu=(0, V^x, V^y, 0).
\end{align}
Note that one has $V_T^2=-|\mathbf{V}|^2$. Here, the light-like vectors are chosen to be
\begin{align}\label{eq:ni}
n_A^\mu=\bar{n}_B^\mu\equiv(1,0,0,1),\qquad n_B^\mu=\bar{n}_A^\mu\equiv(1,0,0,-1),
\end{align}
and the transverse metric is defined as
\begin{align}
    g_T^{\mu\nu} = g^{\mu\nu} - \frac{n_A^\mu n_B^\nu+n_A^\nu n_B^\mu}{2}.
\end{align}
As a result, the integrand on the right-hand side of Eq. (\ref{eq:dPdb}) depends on $\mathbf{\mathcal{P}}_i$ only through the transverse Wigner functions. In order to integrate out the wave packets and get unity, one needs to drop $\mathbf{X}_i$ from its phase factor. Condition ii) is sufficient to justify such an approximation. If one goes back to Eq. (\ref{eq:dPdb_full}), this, equivalently, means that the difference between $\mathbf{p}_i$ and $\bar{\mathbf{p}}_i$ is negligible compared to their average. Obviously, this is also the condition needed to eliminate the ambiguity in defining the impact parameter in Eq. (\ref{eq:b}) due to alternative choices of $\mathbf{x}_i$.

At the end, one can integrate out the transverse Wigner functions in Eq. (\ref{eq:dPdb}) and obtain the impact-parameter dependent differential cross section for producing any observable $O$
\begin{align}\label{eq:dsigmadbdo}
        \frac{d\sigma}{d^2{\mathbf b} dO}  =&\int\frac{d^2\mathbf{q}}{(2\pi)^2}e^{i\mathbf{q}\cdot\mathbf{b}}\int\prod\limits_f\left[d\Gamma_{p_f}\right]\delta(O-O(\{p_f\}))\notag\\
        &\times\frac{1}{2s} M(p_A, p_B \to \{p_f\}) M^*(\bar{p}_A, \bar{p}_B \to \{p_f\})(2\pi)^4\delta^{(4)}(p_A + p_B - \sum p_f),
\end{align}
where $O(\{p_f\})$ defines the observable $O$ as a function of the final-state momenta $\{p_f\}$, the incoming momenta are given in Eq. (\ref{eq:psincoming}) and the phase-space measure in $d$-dimensional spacetime for a particle of mass $m$
\begin{align}\label{eq:Gammapf}
    \int d\Gamma_{p}\equiv \int \frac{d^dp}{(2\pi)^d}(2\pi)\delta(p^2-m^2)\theta(p^0)=\int \frac{dy d^{d-2}\mathbf{p}}{2(2\pi)^{d-1}}
\end{align}
with $y$ the particle's rapidity.

\section{Inclusive hard processes in hadron and nuclear collisions}
\label{sec:nuclear}

Based on the heuristic argument outlined in the introduction, one can expect that a new type of PDFs, which describe the parton distributions in transverse phase space, can be universally defined for inclusive hard processes in impact-parameter dependent hadron and nuclear collisions. In this section, we justify this argument using SCET~\cite{Bauer:2000yr,Bauer:2001ct,Bauer:2001yt,Bauer:2002nz,Beneke:2002ph}. 

\subsection{Factorization for inclusive hard processes with colorless final states}
\label{sec:factorization}
In this subsection, we derive a factorized form of Eq. (\ref{eq:dsigmadbdo}) for the process in Eq. (\ref{eq:process}). In hadron and nuclear collisions, there are additional length scales that one  needs to consider: the sizes $R_i$ of the colliding particles. 
Accordingly, the impact-parameter dependent cross section for hadron and nuclear collisions is defined in the range:\footnote{
We are only interested in strong interactions and hence ignore the cases with $b\gg R_A + R_B$. 
} $R_A + R_B\gtrsim b \gg \Delta x_T$. It always involves  non-perturbative modes with $q_T\sim 1/b \sim R_i$ in addition to perturbative modes with $p_f^2 \gtrsim \Lambda_{QCD}^2$ that contribute to the observable $O(\{p_f\})$ to be measured. Since these non-perturbative modes are collinear to the colliding hadrons or nuclei, we take them as submodes of the corresponding beam collinear modes.

\subsubsection{Basics of SCET}

We first briefly review the elements of SCET that are relevant to our discussion. With the modes of $q_T\sim 1/b$ taken as submodes of the corresponding beam collinear modes, all the relevant infrared degrees of freedom for the process under study are the same as those for the conventional cross section in Refs. \cite{Bauer:2002nz,Mantry:2009qz,Becher:2010tm}:
\begin{align}\label{eq:scalings}
&\text{$n_i$-{\rm collinear:}}~p_{n_i}^\mu \sim  Q \,(\lambda^2,1,\lambda)_{n_i\bar n_i}, \notag \\
&\text{soft: }~p_s^\mu \sim  \left\{\begin{array}{ll}
     \lambda^2 Q &  \qquad\text{for SCET}_{\text{I}}\\
     \lambda Q & \qquad\text{for SCET}_{\text{II}}
\end{array}\right.,
\end{align}
where $\lambda\ll 1$ is an expansion parameter and we have defined for any $V^\mu$
\begin{align}
    V^\mu = \frac{\bar{n}_i^\mu}{2}n_i\cdot V + \frac{n_i^\mu}{2} \bar{n}_i\cdot V + V_T^\mu\equiv(n_i\cdot V, \bar{n}_i\cdot V,\mathbf{V})_{n_i\bar{n}_i}
\end{align}
in terms of a pair of light-like vectors $n_i$ and $\bar{n}_i$ with $n_i\cdot\bar{n}_i=2$.

The $S$-matrix element at leading order in $\lambda$ can be expressed generically in the form
\begin{align}\label{eq:S}
    \langle p_C,\{p_X\} |\hat{S}|\phi_A \phi_B\rangle=\int d^4x e^{ip_C\cdot x}
    \langle \{p_X\}|i\hat{M}(x)|\phi_A \phi_B\rangle
\end{align}
with  $\{p_X\}$ standing for momenta of unmeasured infrared partons and the amplitude operator $\hat{M}$ being a convolution of relevant SCET operators and corresponding Wilson coefficients. Let us combine everything but the collinear and soft fields in the SCET operators with the Wilson coefficients and denote their sum by the coefficient $\mathcal{C}$. In this way, irrespective of the species of the product $C$, $\hat{M}$ can be always written in the form
\begin{align}\label{eq:H}
i\hat{M}(x)=&\int dt_A dt_B \mathcal{C}^{a_Aa_B}_{\alpha_A\alpha_B} (\epsilon, t_A, t_B)[S_{n_A}\phi_{n_A}(x+t_A \bar n_A)]^{\alpha_A}_{a_A}[S_{n_B}\phi_{n_B}(x+t_B \bar n_B)]^{\alpha_B}_{a_B},
\end{align}
where the coefficient $\mathcal{C}$ is to be determined by the matching procedure after the species of the product $C$ being specified, $a_i$ and $\alpha_i$ are respectively the color and Lorentz/spinor indices, $S_{n_i}$ is the soft Wilson line along the collinear direction $n_i$ and the collinear building blocks $\phi_{n_i}$ stand for
\cite{Hill:2002vw}
\begin{align}\label{eq:collinearFields}
\chi_{n_i}(x)=W_{n_i}^\dagger (x) \frac{\slashed{n}_i\slashed{\bar{n}}_i}{4} \psi_{n_i}(x),\qquad\bar\chi_{n_i}(x),\qquad \mathcal{B}_{n_iT}^\mu=\frac{1}{g_s}W_{n_i}^\dagger(x) i D_{n_iT}^\mu W_{n_i}(x),
\end{align}
respectively for $n_i$-collinear quarks, antiquarks or gluons with $D_{n_iT}^\mu\equiv\partial^\mu_T -i g_s A_{n_iT}^\mu$ and $W_{n_i}$ the $n_i$-collinear Wilson line.

Both the hard-collinear factorization and the soft-collinear factorization (at leading order in $\lambda$) are implemented through $\hat{M}$, which are independent of the initial states of the colliding particles. Soft radiation decouples from the hard scattering encoded in the coefficient $\mathcal{C}$, which has been proved for certain processes based on infrared power counting in perturbative QCD~\cite{Collins:1989gx, Sterman:1995fz}. Soft gluons can couple to collinear partons only through their unphysical polarizations $n_i\cdot A_s$, which gives rise to the soft Wilson lines in $\hat{M}$. Soft and collinear fields decouple in the SCET Lagrangian for both SCET$_{\text{I}}$ (after decoupling transformation~\cite{Bauer:2001yt}) and SCET$_{\text{II}}$. The soft Wilson lines in coordinate space take the form\footnote{
This is equivalent to the approximation $(p_c + p_s)^2\approx \bar{n}\cdot p_c n\cdot p_s$ in all propagator denominators with internal momenta given by a sum of $n$-collinear ($p_c$) and soft ($p_s$) momenta. The justification of this approximation in perturbative QCD is rather technical event for the Drell-Yan process~\cite{Collins:1988ig}.
}
\begin{align}
    S_{n_i}(x)=\left\{\begin{array}{ll}
         Pe^{ig_s\int_{-\infty}^0 dt n_i\cdot A_s(tn_i+x)}& \text{for $\phi_{n_i+}$ (incoming particles)} \\
         \bar{P}e^{-ig_s\int^{\infty}_0 dt n_i\cdot A_s(tn_i+x)}& \text{for $\phi_{n_i-}$ (outgoing antiparticles)}
    \end{array}
    \right.
\end{align}
where $\phi_{n_i\pm}$ respectively stand for the positive and negative energy parts of $\phi_{n_i}$ and $A_s\equiv A_s^a T^a$ with $T^a$ being $SU(N_c)$ generators in the corresponding color representation. The collinear Wilson lines take the form
\begin{align}
    W_{n_i}(x) = Pe^{ig_s\int_{-\infty}^0 dt \bar{n}_i\cdot A_{n_i}(t\bar{n}_i+x)}\qquad\text{or}\qquad \bar{P}e^{-ig_s\int_{0}^{\infty} dt \bar{n}_i\cdot A_{n_i}(t\bar{n}_i+x)}.
\end{align}
The inclusion of collinear Wilson lines in the  collinear building blocks is mandated by the collinear gauge invariance~\cite{Bauer:2001yt}.

\subsubsection{Derivation of the factorization formula}
With the $S$-matrix element given in Eq. (\ref{eq:S}), we are ready to derive the factorized cross section for the process in Eq. (\ref{eq:process}). We impose the two conditions in Eq. (\ref{eq:cond}) and identify the impact-parameter dependent probability $P_{\mathbf{b}}$ with the cross section at the outset:
\begin{align}\label{eq:dsigmadb}
\frac{d\sigma}{d^2\mathbf{b}dy_C d^2\mathbf{p}_C} &= \frac{1}{2 (2\pi)^3}\int d^4 X d^4 x~e^{-i p_C\cdot x}
\notag\\
&\times\sum\limits_{\{p_X\}}\langle \phi_A \phi_B|\hat{M}^\dagger\left(X+\frac{x}{2}\right)|\{p_X\}\rangle\langle \{p_X\}|\hat{M}\left(X - \frac{x}{2}\right)|\phi_A \phi_B\rangle.
\end{align}
Here, for definiteness we measure the rapidity $y_C$ and the transverse momentum $\mathbf{p}_C$ of the object $C$.

The following derivation is pretty much the same as that in the previous section, except that we keep $\mathbf{X}$ unintegrated. It tells us the transverse location of the hard scattering vertex as shown in Fig. \ref{fig:sigmb}. One can first integrate out $X^0$ and $X^z$ in Eq. (\ref{eq:dsigmadb}) by using the momentum operator and then the longitudinal momenta associated with the wave packets in the conjugate amplitude. After that, one identifies the longitudinal momenta in the amplitude and the conjugate amplitude, as in Eq. (\ref{eq:cond1}). And, in terms of the transverse Wigner functions in Eq. (\ref{eq:WT}), the cross section can be expressed as
\begin{align}\label{eq:sigma_all}
   \frac{d\sigma}{d^2\mathbf{b} dy_C d^2\mathbf{p}_C} = \int d^2\mathbf{X}&\prod\limits_{i=A,B}\left[\int d^2\mathbf{X}_i d^2\mathbf{\mathcal{P}}_i  \frac{d^2\mathbf{q}_i}{(2\pi)^2}e^{i\mathbf{q}_i\cdot\left(\mathbf{X}-\mathbf{x}_i - \mathbf{X}_i \right)}W_i(\mathbf{X}_i, \mathbf{\mathcal{P}}_i)\right]\notag\\
    &\times \frac{d\sigma}{dy_C d^2\mathbf{p}_C}(p_A, p_B\to p_C \leftarrow\bar{p}_A, \bar{p}_B),
\end{align}
where in terms of $\hat{M}$, the off-diagonal cross section in Eq. (\ref{eq:offsigmaLO}) takes the form 
\begin{align}\label{eq:offsigmaSCET}
    \frac{d\sigma}{dy_C d^2\mathbf{p}_C}&(p_A, p_B\to p_C \leftarrow\bar{p}_A, \bar{p}_B)
    =\frac{1}{2 (2\pi)^3}\frac{1}{2s}\int d^4 x e^{-i p_C\cdot x}\notag\\
    &\times\sum\limits_{\{p_X\}}\langle \bar{p}_A \bar{p}_B|\hat{M}^\dagger\left(\frac{x}{2}\right)|\{p_X\}\rangle\langle \{p_X\}|\hat{M}\left(- \frac{x}{2}\right)|p_A p_B\rangle,
\end{align}
and the incoming momenta are given by
\begin{align}\label{eq:psincomingSCET}
    &p_A^\mu = \bar{n}_A\cdot P_A\frac{n_A^\mu}{2}+ \frac{q_A^\mu}{2}- \frac{q_A^2}{4 \bar{n}_A\cdot P_A} \frac{\bar{n}_A^\mu}{2},\qquad p_B^\mu = \bar{n}_B\cdot P_B\frac{n_B^\mu}{2}+ \frac{q_{B}^\mu}{2}- \frac{q_B^2}{4 \bar{n}_B\cdot P_B} \frac{\bar{n}_B^\mu}{2},\notag\\
    &\bar{p}_A^\mu = \bar{n}_A\cdot P_A\frac{n_A^\mu}{2}- \frac{q_A^\mu}{2}- \frac{q_A^2}{4 \bar{n}_A\cdot P_A} \frac{\bar{n}_A^\mu}{2},\qquad \bar{p}_B^\mu =\bar{n}_B\cdot P_B\frac{n_B^\mu}{2}- \frac{q_B^\mu}{2}- \frac{q_B^2}{4 \bar{n}_B\cdot P_B} \frac{\bar{n}_B^\mu}{2},
\end{align}
with $q_i^\mu= p_i^\mu-\bar{p}_i^\mu$ orthogonal to $n_i$ and $\bar{n}_i$. Finally, the two conditions in Eq. (\ref{eq:cond}) allow one to integrate out the transverse Wigner functions and one has
\begin{align}\label{eq:dsigmadbSCET}
   \frac{d\sigma}{d^2\mathbf{b} dy_C d^2\mathbf{p}_C} = &\int d^2\mathbf{X}\int\limits_{\mathbf{q}_A, \mathbf{q}_B} e^{i  \mathbf{q}_A\cdot(\mathbf{X}-\mathbf{x}_A)+i \mathbf{q}_B\cdot(\mathbf{X}-\mathbf{x}_B)} \frac{d\sigma}{dy_C d^2\mathbf{p}_C}(p_A, p_B\to p_C \leftarrow\bar{p}_A, \bar{p}_B).
\end{align}

Since the collinear and soft modes decouple, the right-hand side of the above equation can be further written in a factorized form. The coefficient $\mathcal{C}$ in $\hat{M}$ is independent of the initial states of the colliding particles, which is combined into the hard function together with its complex conjugate. The soft Wilson lines only act on the vacuum state to define the soft function. As a result, the soft function is independent of $\mathbf{X}$, as shown below. That is, the soft and hard functions are the same as for the conventional cross section~\cite{Bauer:2002nz,Mantry:2009qz,Becher:2010tm}.

The collinear fields act on the incoming states to give a new type of PDFs, which are referred to as {\it thickness beam functions} in this paper. The $n_i$-collinear sector in Eq. (\ref{eq:dsigmadbSCET}) takes the form
\begin{align}\label{eq:W}
  \mathcal{T}_{a'a}^{\alpha'\alpha}&(\mathbf{r}_i, n_i\cdot x, \mathbf{x})=\int\frac{d^2\mathbf{q}}{(2\pi)^2}e^{i \mathbf{q}\cdot\mathbf{r}_i}\notag\\
  &\times\left\langle \bar{n}_i\cdot P_i, -\frac{\mathbf{q}}{2} \right|[\phi_{n_i}^\dagger]^{\alpha'}_{a'}\left(\frac{n_i\cdot x}{2},\frac{\mathbf{x}}{2}\right)\left[\phi_{n_i}\right]^{\alpha}_{a}\left(- \frac{n_i\cdot x}{2},-\frac{\mathbf{x}}{2}\right)\left|\bar{n}_i\cdot P_i, \frac{\mathbf{q}}{2}\right\rangle
\end{align}
with $\mathbf{r}_i\equiv\mathbf{X}-\mathbf{x}_i$ and the spin of particle $i$ being implicitly averaged over. Given some projector $P^{\alpha' \alpha}$, we define the corresponding thickness beam function $\mathcal{T}_{j/i}$ as
\begin{align}\label{eq:fb}
\mathcal{T}_{a'a}^{\alpha'\alpha}(\mathbf{r}_i, {n_i}\cdot x, \mathbf{x})\to \int \frac{dz}{z}\frac{P^{\alpha' \alpha}(z\bar{n}_i\cdot P_i)}{d_{c_i}}\delta_{a'a} \mathcal{T}_{j/i}(\mathbf{r}_i, z, \mathbf{x})e^{i\frac{n_i\cdot x}{2}z\bar{n}_i\cdot P_i}
\end{align}
with $d_{c_i}$ the dimension of the color representation of $\phi_{n_i}$ and $j$ the parton species corresponding to $\phi_{n_i}$. The most common projectors for unpolarized collisions are given by the spin/polarization average, which take the form
\begin{align}\label{eq:Ppol}
    P^{\bar\alpha_i \alpha_i}(k)=\left\{
    \begin{array}{ll}
         \frac{1}{2}\left(\slashed{k}\right)^{\bar{\alpha}_i\alpha_i} & \text{for quarks/antiquarks} \\
         \frac{1}{d-2}(-g_T^{\bar{\alpha}_i\alpha_i})& \text{for gluons}
    \end{array}
    \right.
\end{align}
with $d$ the spacetime dimension.

Inserting the expression of $\hat{M}$ in Eq. (\ref{eq:H}) into Eq. (\ref{eq:dsigmadbSCET}) and
making the replacement in Eq. (\ref{eq:fb}) eventually yields the factorization formula:
\begin{align}\label{eq:factorization}
   \frac{d\sigma_{AB}}{d^2\mathbf{b} dy_C d^2\mathbf{p}_C} 
     =& \frac{1}{4\pi s}\sum\limits_{j,k}\int d^2\mathbf{X}\int d^2\mathbf{x}e^{i\mathbf{p}_C\cdot\mathbf{x}}\int_0^1\frac{dz_A}{z_A}\frac{dz_B}{z_B}\mathcal{T}_{j/A}(\mathbf{X}, z_A,\mathbf{x}) \mathcal{T}_{k/B}(\mathbf{X} -\mathbf{b}, z_B,\mathbf{x})\notag\\
     &\times \int\prod\limits_f [d\Gamma_{p_f}]\prod\limits_{i=A,B}\delta(z_i \bar{n}_i \cdot P_i- \bar{n}_i\cdot p_C- \sum\bar{n}_i\cdot p_f)\notag\\
&\times H_{a_A a_B}^{\bar{a}_A \bar{a}_B}\left(z_A P_A, z_B P_B\to p_C, \{p_f\}\right)\mathcal{S}_{a_A a_B}^{\bar{a}_A \bar{a}_B}(\mathbf{x}),
\end{align}
where we have taken $\mathbf{x}_A=0$ and $\mathbf{x}_B=\mathbf{b}$, $P_i^\mu=\frac{\bar{n}_i\cdot P_i}{2}n_i^\mu$, the soft function is defined by
\begin{align}\label{eq:softFun}
    \mathcal{S}_{a_A a_B}^{\bar{a}_A \bar{a}_B}(\mathbf{x})&\equiv\langle 0|\bar{T}[S_{n_B}^{\dagger a'_B \bar{a}_B}(\mathbf{x}_+)S_{n_A}^{\dagger a'_A \bar{a}_A}(\mathbf{x}_+)]T[S^{a_A a'_A}_{n_A}(\mathbf{x}_-)S^{a_B a'_B}_{n_B}(\mathbf{x}_-)]|0 \rangle\notag\\
    &=\langle 0|\bar{T}[S_{n_B}^{\dagger a'_B \bar{a}_B}(\mathbf{x})S_{n_A}^{\dagger a'_A \bar{a}_A}(\mathbf{x})]T[S^{a_A a'_A}_{n_A}(0)S^{a_B a'_B}_{n_B}(0)]|0 \rangle
\end{align}
with $\mathbf{x}_\pm\equiv \mathbf{X}\pm \mathbf{x}/2$, and the hard function, given by the partonic process $j (z_A P_A)+k (z_B P_B)\to C(p_C)$ + anything else$(\{p_f\})$, takes the form
\begin{align}\label{eq:hard}
    H_{a_A a_B}^{\bar{a}_A \bar{a}_B}\equiv \frac{P^{\bar\alpha_A \alpha_A}}{d_{c_A}}\frac{P^{\bar\alpha_B \alpha_B}}{d_{c_B}}\mathcal{\tilde{C}}^{*\bar{a}_A\bar{a}_B}_{\bar{\alpha}_A\bar{\alpha}_B}\mathcal{\tilde{C}}^{a_Aa_B}_{\alpha_A\alpha_B}
\end{align}
with $\mathcal{\tilde{C}}$ given by
\begin{align}
    \mathcal{\tilde{C}}(\epsilon,z_A\bar{n}_A\cdot P_A, z_B\bar{n}_B\cdot P_B)=\int dt_A dt_B e^{i (t_A z_A\bar{n}_A\cdot P_A+t_B z_B\bar{n}_B\cdot P_B)}\mathcal{C} (\epsilon, t_A, t_B).
\end{align}
For the spin-averaged projectors in Eq. (\ref{eq:Ppol}), the hard function
\begin{align}\label{eq:hardM2}
H_{a_A a_B}^{{a}_A {a}_B}=\overline{|M|}^2
\end{align}
with $\overline{|M|}^2$ the square of the amplitude averaged over initial-state colors, spins or polarizations~\cite{Becher:2014oda}.

\subsection{Thickness beam functions and transverse phase-space parton distributions}
\label{sec:T}
The thickness beam functions are universal. Their emergence only relies on the hard-collinear and soft-collinear factorization. Therefore, they should universally show up in all inclusive hard processes in hadron and nuclear collisions as long as the processes admit of these two types of factorization. And hard processes with colorless final states like that in Eq. (\ref{eq:process}) provide the cleanest way to measure the thickness beam functions. The discussion in the previous subsection can be easily generalized to deep inelastic scattering in electron-proton and electron-ion collisions. Therefore, they could also be measured using the future Electron-Ion Collider \cite{Accardi:2012qut} once the impact-parameter dependence of the collisions can be determined experimentally.

The thickness beam functions are related to transverse phase-space PDFs (TPS PDFs)\footnote{
TPS PDFs have a corresponding definition in QCD, in which the collinear fields are replaced by the corresponding parton fields and the collinear Wilson lines are combined into gauge links. Subtlety, however, could arise from the difference between the product of collinear Wilson lines and the chosen gauge links as for the beam functions~\cite{Stewart:2009yx} and TMD PDFs~\cite{Collins:1981uw}. 
} via the Fourier transform with respect to $\mathbf{x}$:
\begin{align}\label{eq:tpspdfs}
    f_{j/i}(\mathbf{r}, z, \mathbf{p})=\int d^2\mathbf{x} e^{i \mathbf{p}\cdot\mathbf{x}}\mathcal{T}_{j/i}(\mathbf{r}, z, \mathbf{x}).
\end{align}
They are
two-dimensional analogues of quantum phase-space distributions in the rest frame of a proton defined in Ref. \cite{Belitsky:2003nz}. This can be easily seen from the definition of $\mathcal{T}_{j/i}$ in Eq. (\ref{eq:fb}), which generally takes the form
\begin{align}
    \mathcal{T}_{j/i}(\mathbf{r}, z, \mathbf{x})=&\int\frac{d^2\mathbf{q}}{(2\pi)^2}e^{i\mathbf{q}\cdot\mathbf{r}}\int\frac{dt}{2\pi}e^{-i z t \bar{n}\cdot P}\notag\\
  &\times\left\langle \bar{n}\cdot P, -\frac{\mathbf{q}}{2} \right|[\phi_{n}^\dagger]^{\alpha'}_{a}\left(\frac{t \bar{n}}{2}+ \frac{x_T}{2}\right)\Gamma_{\alpha'\alpha}[\phi_{n}]^{\alpha}_{a}\left(-\frac{t \bar{n}}{2}- \frac{x_T}{2}\right)\left|\bar{n}\cdot P, \frac{\mathbf{q}}{2}\right\rangle
\end{align}
with $\Gamma_{\alpha'\alpha}$ to be determined after $P^{\alpha'\alpha}$ are chosen. By using the momentum operator, one can write
\begin{align}\label{eq:Tdef}
    \mathcal{T}_{j/i}(\mathbf{r}, z, \mathbf{x})=&\sum\limits_m\int\frac{d^2\mathbf{q}}{(2\pi)^2}\left\langle \bar{n}\cdot P, -\frac{\mathbf{q}}{2} |[\phi_{n}^\dagger]^{\alpha'}_{a}\left( \mathbf{r}\right)|m\right\rangle\Gamma_{\alpha'\alpha}\left\langle
   m |[\phi_{n}]^{\alpha}_{a}\left( \mathbf{r}\right)|\bar{n}\cdot P, \frac{\mathbf{q}}{2}\right\rangle\notag\\
  &\times e^{i\mathbf{p}_m\cdot \mathbf{x}}\delta(\bar{n}\cdot p_m-(1-z)\bar{n}\cdot P).
\end{align}
In the first line, the right-hand side of this equation can be interpreted as the ``probability"\footnote{Like the Wigner function in quantum mechanics \cite{Wigner:1932eb}, we don't expect that this term is always positive. It can be literally interpreted as a probability only after being coarse-grained, say, measured at $\mathbf{r}$ with a relatively large uncertainty.
}
of measuring a collinear field of type $j$ at
$\mathbf{r}$ inside particle $i$ and producing a final state $|m\rangle$ with total momentum $p_m^\mu$. Accordingly, the corresponding TPS PDF is
\begin{align}
    f_{j/i}(\mathbf{r}, z, \mathbf{p})=&\sum\limits_m\int\frac{d^2\mathbf{q}}{(2\pi)^2}\left\langle \bar{n}\cdot P, -\frac{\mathbf{q}}{2} |[\phi_{n}^\dagger]^{\alpha'}_{a}\left( \mathbf{r}\right)|m\right\rangle\Gamma_{\alpha'\alpha}\left\langle
   m |[\phi_{n}]^{\alpha}_{a}\left( \mathbf{r}\right)|\bar{n}\cdot P, \frac{\mathbf{q}}{2}\right\rangle\notag\\
  &\times \delta^{(2)}(\mathbf{p}+\mathbf{p_m})\delta(\bar{n}\cdot p_m-(1-z)\bar{n}\cdot P),
\end{align}
which, after being coarse-grained, is the probability of finding a collinear parton at
$\mathbf{r}$ carrying a momentum fraction $z$ and transverse momentum $\mathbf{p}$.

In the aforementioned coarse-grained sense, thickness beam functions and TPS PDFs admit interpretation respectively as the beam functions~\cite{Stewart:2009yx} and TMD PDFs~\cite{Collins:1981uw} at $\mathbf{r}$ since one typically has $|\mathbf{r}|\gg |\mathbf{x}|$ for hard processes. For inclusive hard processes with $|\mathbf{x}|\sim 1/Q$, $\mathbf{x}$ can be dropped from the thickness beam functions as a result of multipole expansion. And $\mathcal{T}_{j/i}(\mathbf{r}, z, \mathbf{0})$ can be viewed as the corresponding conventional parton distribution functions~\cite{Curci:1980uw, Collins:1981uw} at $\mathbf{r}$, that is, the transverse spatial PDFs. In this case, the off-diagonal matrix element needed for defining $\mathcal{T}_{j/i}(\mathbf{r}, z, \mathbf{0})$ is the same as generalized parton distributions (GPDs)~\cite{Diehl:2003ny} with the incoming momenta differing only in the transverse plane.

At the end, we explicitly write down the expressions for the thickness beam functions corresponding to the spin-averaged projectors in Eq. (\ref{eq:Ppol}):

\noindent
for quarks,
 \begin{align}\label{eq:Wq}
  \mathcal{T}_{q/i}(\mathbf{r}, z,\mathbf{x})=&\int\frac{d^2\mathbf{q}}{(2\pi)^2}e^{i\mathbf{q}\cdot\mathbf{r}}\int\frac{dt}{2\pi}e^{-i z t \bar{n}\cdot P}\notag\\
  &\times\left\langle \bar{n}\cdot P, -\frac{\mathbf{q}}{2} \right|\bar\chi_{n}\left(\frac{t \bar{n}}{2}+ \frac{x_T}{2}\right)\frac{\slashed{\bar{n}}}{2}\chi_{n}\left(-\frac{t \bar{n}}{2}- \frac{x_T}{2}\right)\left|\bar{n}\cdot P, \frac{\mathbf{q}}{2}\right\rangle,
\end{align}
for antiquarks,
\begin{align}\label{eq:Wqb}
  \mathcal{T}_{\bar{q}/i}(\mathbf{r}, z,\mathbf{x})=&\int\frac{d^2\mathbf{q}}{(2\pi)^2}e^{i\mathbf{q}\cdot\mathbf{r}}\int\frac{dt}{2\pi}e^{-i z t \bar{n}\cdot P}\notag\\
  &\times\left\langle \bar{n}\cdot P, -\frac{\mathbf{q}}{2} \right|\text{Tr}\left[\frac{\slashed{\bar{n}}}{2}\chi_{n}\left(\frac{t \bar{n}}{2}+ \frac{x_T}{2}\right)\bar\chi_{n}\left(-\frac{t \bar{n}}{2}- \frac{x_T}{2}\right)\right]\left|\bar{n}\cdot P, \frac{\mathbf{q}}{2}\right\rangle\notag\\
  =&\int\frac{d^2\mathbf{q}}{(2\pi)^2}e^{i\mathbf{q}\cdot\mathbf{r}}\int\frac{dt}{2\pi}e^{-i z t \bar{n}\cdot P}\notag\\
  &\times(-)\left\langle \bar{n}\cdot P, -\frac{\mathbf{q}}{2} \right|\bar\chi_{n}\left(-\frac{t \bar{n}}{2}- \frac{x_T}{2}\right)\frac{\slashed{\bar{n}}}{2}\chi_{n}\left(\frac{t \bar{n}}{2}+ \frac{x_T}{2}\right)\left|\bar{n}\cdot P, \frac{\mathbf{q}}{2}\right\rangle,
\end{align}
and for gluons
 \begin{align}\label{eq:Wg}
  \mathcal{T}_{g/i}(\mathbf{r}, z,\mathbf{x})=&z\bar{n}\cdot P(-{g_T}_{\alpha'\alpha})\int\frac{d^2\mathbf{q}}{(2\pi)^2}e^{i\mathbf{q}\cdot\mathbf{r}}\int\frac{dt}{2\pi}e^{-i z t \bar{n}\cdot P}\notag\\
  &\times\left\langle \bar{n}\cdot P, -\frac{\mathbf{q}}{2} \right|\mathcal{B}_{nT}^{a\alpha'}\left(\frac{t \bar{n}}{2}+ \frac{x_T}{2}\right)\mathcal{B}_{nT}^{a\alpha}\left(-\frac{t \bar{n}}{2}- \frac{x_T}{2}\right)\left|\bar{n}\cdot P, \frac{\mathbf{q}}{2}\right\rangle.
\end{align}
For gluons, there are also other possible projectors as combinations of $g_T^{\mu\nu}$, $x_T^\mu$ and $r_T^\mu$. We will not exhaust all the possible forms of the projectors in this paper, whose relevance will depend on specific processes under consideration. 
\subsection{The Glauber model for hard processes in heavy-ion collisions}
\label{sec:glauber}
In heavy-ion collisions, one has 
\begin{align}
|\mathbf{r}_i| \sim |\mathbf{b}|\sim R_i\gg 1/\Lambda_{QCD}\gg |\mathbf{x}|, 1/Q.
\end{align}
That is, $\mathbf{r}_i$-dependence of thickness beam functions lies deep in the non-perturbative regime. In principle, one could also quantitatively study such non-perturbative degrees of freedom using the effective field theory approach. We, instead, content ourselves with connecting our factorization formula in Eq. (\ref{eq:factorization}) to the cross section in the Glauber model~\cite{Miller:2007ri} by modelling large nuclei.

In heavy-ion collisions, the two conditions in Eq. (\ref{eq:cond}) are easily fulfilled. In principle, the impact parameter in Eq. (\ref{eq:b}) can be determined better than 1 fm
as long as one is not aimed at high accuracy in the determination of the beam particles' transverse momenta and keeps $\Delta p_T\gtrsim$ 1 GeV\footnote{
In  this paper, we are not concerned about how to determine the impact parameter of a collision experimentally but only about what is speakable and unspeakable about the impact parameter in the quantum picture. 
}. Since the quantum fuzziness is much smaller than the impact parameter, the classical concept of collision geometry used in the Glauber model is indeed a reasonable approximation to the underlying quantum picture.

Let us approximate the thickness beam functions by appealing to a commonly used model for heavy nuclei as in the Glauber model~\cite{Miller:2007ri}. We also consider the difference between protons and neutrons~\cite{Helenius:2012wd}. Large nuclei are known to be loosely bound with the binding energy per nucleon $\Delta E\approx 8$ MeV in their rest frame. This means that the typical time scale for internal nucleon interactions is of order $\Delta t = \frac{\bar{n}\cdot P_i}{2 m_i}\frac{1}{\Delta E}\approx25\gamma$ fm in the lab frame. Such a time scale is much longer than any other time scales in the problem. Therefore, nucleons in the nuclei are to be treated as free particles with a normalized distribution $\hat{\rho}_i$ proportional to that of electric charges~\cite{DeJager:1987qc}. Accordingly, the probability to find a nucleon per unit transverse area around $\mathbf{r}_i$ is given by
\begin{align}
    \hat{T}_i(\mathbf{r}_i)\equiv\int dz \hat{\rho}_i(\mathbf{r}_i, z).
\end{align}
And, the thickness beam functions of nucleus $i$, which is made of $Z_i$ protons and $N_i$ neutrons, can be replaced by
\begin{align}\label{eq:Tanzart}
    \mathcal{T}_{j/i}(\mathbf{r}_i,z, \mathbf{x})\to T_i(\mathbf{r}_i)\left[\frac{Z_i}{Z_i+N_i}B_{j/p}(z,\mathbf{x}) + \frac{N_i}{Z_i+N_i}B_{j/n}(z,\mathbf{x})\right],
\end{align}
where the nuclear thickness function is defined as~\cite{Miller:2007ri}
\begin{align}
    T_i(\mathbf{r}_i)\equiv(Z_i + N_i)\hat{T}_i(\mathbf{r}_i),
\end{align} 
$B_{j/p}$ and $B_{j/n}$ are the beam functions for protons and neutrons, respectively. That is, in the Glauber model, the thickness beam functions are products of the thickness functions and the beam functions.

Inserting Eq. (\ref{eq:Tanzart}) into Eq. (\ref{eq:factorization}) gives the cross section for hard processes in the Glauber model
\begin{align}\label{eq:sigmaGlauber}
   \frac{d\sigma_{AB}}{d^2\mathbf{b} dy_C d^2\mathbf{p}_C} 
     = \int d^2\mathbf{X}T_{A}(\mathbf{X}) T_{B}(\mathbf{X} -\mathbf{b})\frac{d\sigma_{nn}}{dy_C d^2\mathbf{p}_C},
\end{align}
with the binary nucleon-nucleon cross section $\sigma_{nn}$ given by the factorization formula for colliding two nucleon beams with neutron-to-proton ratios respectively equal to $N_i/Z_i$. That is, the nuclear modification factor\footnote{
In experiments, one measures $R_{AA}$ with both its numerator and denominator averaged over a range of impact parameter $\Delta b$  corresponding to some centrality bin. Since $\Delta b\gg1/|\mathbf{p}_C|$, one can identify the average impact-parameter dependent cross section with the average number of hard processes per collision.
}
\begin{align}\label{eq:RAA}
    R_{AA}\equiv\frac{\frac{d\sigma_{AB}}{{d^2\mathbf{b}dy_C d^2\mathbf{p}_C}}}{T_{AB}(\mathbf{b}) \frac{d\sigma_{nn}}{dy_C d^2\mathbf{p}_C}}=1,
\end{align}
with
\begin{align}
    T_{AB}(\mathbf{b})\equiv \int d^2\mathbf{X}T_{A}(\mathbf{X}) T_{B}(\mathbf{X} -\mathbf{b}).
\end{align}

\subsection{Impact-parameter dependent pp collisions}
\label{sec:pp}
Collective phenomena among produced soft particles have been studied in pp collisions~\cite{Khachatryan:2010gv,Aad:2015gqa,Khachatryan:2015lva, Khachatryan:2016txc,Aaboud:2017blb,Sirunyan:2017uyl}, which are presumptively related to collision geometry according to some models~\cite{Nagle:2018nvi}. Inclusive hard processes in impact-parameter dependent pp collisions are worth being explored experimentally as well. Based on the factorization formula in Eq. (\ref{eq:factorization}), such hard processes can be potentially used to measure transverse phase-space parton distributions inside protons.

Caution is, however, needed when one studies impact-parameter dependent pp collisions. For hard processes, one has 
\begin{align}
|\mathbf{r}_i| \sim |\mathbf{b}|\sim R_i\sim 1/\Lambda_{QCD}\gg |\mathbf{x}|, 1/Q.
\end{align}
Therefore, like heavy-ion collisions, $\mathcal{T}_{j/i}(\mathbf{r}, z, \mathbf{x})$ can be viewed as the distribution of parton $j$ with a transverse size $\sim |\mathbf{x}|$, i.e., the beam function $B_{j/i}(z, \mathbf{x})$, located at $\mathbf{r}$ inside proton $i$. On the  other hand, in contrast to heavy-ion collisions, the two conditions in Eq. (\ref{eq:cond}) are not always fulfilled in pp collisions. As discussed in Sec. \ref{sec:cross_section}, in order to measure universal quantities across experiments, one needs to maintain
\begin{align}
    \Delta p_T\geq \frac{1}{2\Delta x_T}\gg \frac{\Lambda_{QCD}}{2}\approx 100~\text{MeV}.
\end{align}
$\Delta p_T$, however, can not be too large on modern colliders. For example, the crossing angle at the interaction point $\theta_C\sim 100~\mu$rad at the LHC~\cite{Evans:2008zzb}, which is one of the crucial parameters for achieving high luminosity. $\theta_C$ can be determined with an accuracy $\Delta \theta_C=10~\mu$rad, which gives us
\begin{align}
    \Delta p_T \leq 70~\text{MeV}
 \end{align}
for $E=7$ TeV proton beams. If the measurements like those in Refs. \cite{Khachatryan:2010gv,Aad:2015gqa,Khachatryan:2015lva, Khachatryan:2016txc,Aaboud:2017blb,Sirunyan:2017uyl} are subject to a similar constraint (with lower beam energies), the required theoretical calculations always involve the wave packets of colliding protons, as shown in Eqs. (\ref{eq:dPdb}) and (\ref{eq:sigma_all}), which, however, are not measured\footnote{
When Condition ii) in Eq. (\ref{eq:cond}) is violated, the factorization formula in Eq. (\ref{eq:factorization}) is still valid but the definition of thickness beam functions will depend on the protons' wave packets.
}.

\section{The impact-parameter dependent Drell-Yan process in $q\bar{q}$ collisions}
\label{sec:qqb}
The modes associated with $q_T\sim 1/b$ are non-perturbative in hadron and nuclear collisions. In order to make a model-independent verification of the factorization formula in Eq. (\ref{eq:factorization}), in this section we study impact-parameter dependent $q\bar{q}$ collisions with $b\lesssim 1/\Lambda_{QCD}$. Since the factorization formula is valid at all orders in $\alpha_s$ and at leading order in $\lambda$, its validity in such collisions can be verified order by order in perturbation theory by comparing to the results of the general formula in Eq. (\ref{eq:dsigmadbdo}). For this task, the factorization formula is required to reproduce the perturbative QCD results, expanded to leading order in $\lambda$.

\subsection{The impact-parameter dependent Drell-Yan cross section}

We calculate the impact-parameter dependent cross section for the Drell-Yan process
\begin{align}
q\bar{q}\to \gamma^* + \text{anything else}
\end{align}
with the virtuality of the photon $p_C^2 = Q^2$ at the impact parameter $b\lesssim 1/\Lambda_{QCD}$. The  quark and antiquark are taken as massless onshell particles. In this case, there are only two scales: $Q$ and $1/b$, which set the expansion parameter $\lambda=1/(b Q)\ll 1$. Like the conventional cross section, the impact-parameter dependent cross section is not infrared safe because of insufficient average over the initial states. Its singularities are to be regularized by dimensional regularization with $d=4-2\epsilon$.

Let us first derive the  impact-parameter dependent total cross section from the general formula in Eq. (\ref{eq:dsigmadbdo}). One first integrates over the observable $O$ and then singles out one of the final-state particles as $\gamma^*$ to obtain the general formula for the total cross section
\begin{align}\label{eq:sigmaQCDun}
   \frac{d\sigma_{q\bar{q}}}{d^2\mathbf{b}}   =&\frac{\pi}{s}\int\frac{d^2\mathbf{q}}{(2\pi)^2}e^{i\mathbf{q}\cdot\mathbf{b}}\int\prod\limits_f\left[d\Gamma_{p_f}\right]\delta(p_C^2-Q^2)\notag\\
   &\times M(p_A, p_B \to p_C,\{ p_f\}) M^*(\bar{p}_A, \bar{p}_B \to p_C, \{p_f\})
\end{align}
with $p_C=p_A+p_B-\sum p_f$ and the incoming momenta given by Eq. (\ref{eq:psincoming}).
For unpolarized collisions, it can be written in the following compact form
\begin{align}\label{eq:sigmaQCD}
   \frac{d\sigma_{q\bar{q}}}{d^2\mathbf{b}}   =&\frac{\pi}{s}\int\frac{d^2\mathbf{q}}{(2\pi)^2}e^{i\mathbf{q}\cdot\mathbf{b}}\int\prod\limits_f\left[d\Gamma_{p_f}\right]\delta(p_C^2-Q^2) \overline{|M|^2}(p_A, p_B \to p_C, \{p_f\}\leftarrow\bar{p}_A, \bar{p}_B)
\end{align}
with $\overline{|M|^2}(p_A, p_B \to p_C, \{p_f\}\leftarrow\bar{p}_A, \bar{p}_B)$ being the off-diagonal amplitude squared in the last line of Eq. (\ref{eq:sigmaQCDun}) averaged over initial-state spins.

For the factorization formula, one needs to integrate over $y_C$ and $\mathbf{p}_C$ in Eq. (\ref{eq:factorization}). This can be easily done by using Eq. (\ref{eq:Gammapf}) to introduce $\delta(p_C^2-Q^2)$ and then integrate out the four-momentum $p_C^\mu$ instead. We only consider the production threshold: $Q^2\sim s$. In this case, one can replace $\delta(p_C^2-Q^2)$ 
by $\delta(\bar n_A\cdot p_C \bar n_B\cdot p_C-Q^2)$ and the $\mathbf{p}_C$-integral gives $\delta^{(2)}(\mathbf{x})$. Then, integrating out $\mathbf{x}$ gives
\begin{align}\label{eq:factorizationQ}
   \frac{d\sigma_{q\bar{q}}}{d^2\mathbf{b}} 
     =& \frac{\pi}{ s}\sum\limits_{j,k}\int\frac{dz_A}{z_A} \frac{dz_B}{z_B}\int d^2\mathbf{X}\mathcal{T}_{j/q}(\mathbf{X}, z_A) \mathcal{T}_{k/\bar{q}}(\mathbf{X} -\mathbf{b}, z_B)\int\prod\limits_f [d\Gamma_{p_f}]\notag\\
     &\times H\left(z_A P_A, z_B P_B\to p_C, \{p_f\}\right)\delta(\bar n_A\cdot p_C \bar n_B\cdot p_C-Q^2)
\end{align}
with $\bar{n}_i\cdot p_C=z_i \bar{n}_i\cdot P_i-\sum \bar{n}_i\cdot p_f$. Here, we have used the fact that the soft function becomes unity at $\mathbf{x}=0$ and the hard function $H$ is equal to $\overline{|M|^2}$ for the partonic process $j,k\to C$ + anything else, as given in Eq. (\ref{eq:hardM2}). We define
\begin{align}
    \mathcal{T}_{j/i}(\mathbf{r}, z)\equiv \mathcal{T}_{j/i}(\mathbf{r}, z, \mathbf{0}),
\end{align}
which are referred to as transverse spatial PDFs. In heavy-ion collisions, the above equation gives the factorization formula for the inclusive Drell-Yan production of $\gamma^*$, $W^\pm$ and $Z^0$ with a single hard scale $Q$, valid for all orders in $\alpha_s$. The interested reader is referred to Refs. \cite{Vogt:2000hp,Helenius:2012wd} for fixed-order calculations and phenomenological studies. Below, we only focus on the verification of the factorization formula by a detailed calculation for $q\bar{q}$ collisions at leading order in $\lambda$.
 

\subsection{Factorization at the Born level}
At zeroth order in $\alpha_s$, the off-diagonal amplitude squared is given by
\begin{align}\label{eq:M02off}
    \overline{|M^{(0)}|^2}(p_A, p_B\to p_C \leftarrow\bar{p}_A, \bar{p}_B)&=\begin{array}{c}
         \includegraphics[width=0.3\textwidth]{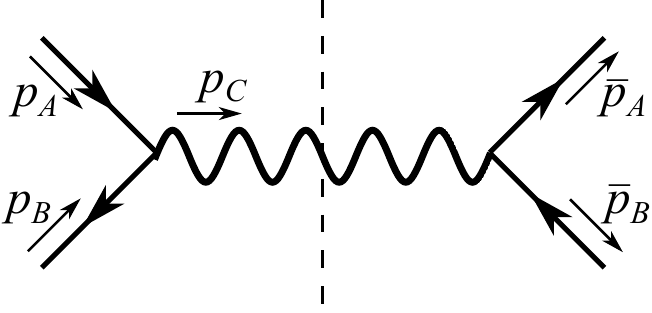}
    \end{array}\notag\\
    &=-\frac{1}{4}e_q^2\text{Tr}\left[v^s(\bar{p}_B) \bar{v}^s(p_B)\gamma^\mu u^{s'}(p_A) \bar{u}^{s'}(\bar{p}_A)\gamma_\mu\right]
\end{align}
with $e_q$ the electric charge of the quark and antiquark. Expand it around $\lambda=0$, and at leading order one has
\begin{align}\label{eq:M2LO}
    \overline{|M^{(0)}|^2}(p_A, p_B\to p_C \leftarrow\bar{p}_A, \bar{p}_B)=e_q^2(1-\epsilon) s,
\end{align}
where $s=\bar n_A\cdot P_A \bar n_B \cdot P_B$ and we have used the following identities
\begin{align}
    &\sum\limits_s u^s(p)\bar{u}^s(p')=\frac{\slashed{p}+m}{\sqrt{2(m+E)}}\left(1+\gamma^0\right)\frac{\slashed{p}'+m}{\sqrt{2(m+E')}},\notag\\
    &\sum\limits_s v^s(p)\bar{v}^s(p')=\frac{-\slashed{p}+m}{\sqrt{2(m+E)}}\left(-1+\gamma^0\right)\frac{-\slashed{p}'+m}{\sqrt{2(m+E')}}.
\end{align}
Inserting Eq. (\ref{eq:M2LO}) into Eq. (\ref{eq:sigmaQCD}) and expanding the $\delta$ function in $Q$ as well\footnote{
As an exercise to illustrate the validity of such an expansion, one can work out
\begin{align}
    Q^2\int\frac{d^2\mathbf{q}}{(2\pi)^2}e^{i\mathbf{q}\cdot\mathbf{b}}\delta(|\mathbf{q}|^2-Q^2 x)=\frac{Q^2}{4\pi}J_0(Q b \sqrt{x})\to \delta^{(2)}(\mathbf{b})\delta(x)\text{ as } Q\to\infty,
\end{align}
by using test functions in both $\mathbf{b}$ and $x$.
} gives
\begin{align}\label{eq:sigm0}
    \frac{d\sigma^{(0)}}{d^2\mathbf{b}} = \pi^2 e_q^2 \delta(s-Q^2)\delta^{(2)}(\mathbf{b}).
\end{align}
That is, the hard scattering is initiated by the quark and antiquark only when they pass very close to one another at a distance $\sim 1/Q$, which defines "point-like" in our calculation. 

Now, let us calculate the factorized Born cross section using  Eq. (\ref{eq:factorizationQ}). At zeroth order in $\alpha_s$, the transverse spatial quark/antiquark distributions, according to Eqs. (\ref{eq:Wq}) and (\ref{eq:Wqb}), take the form
\begin{align}\label{eq:T0qq}
    \mathcal{T}^{(0)}_{q/q}(\mathbf{r}, z)=\mathcal{T}^{(0)}_{\bar{q}/\bar{q}}(\mathbf{r}, z)=\delta(1-z)\delta^{(2)}(\mathbf{r}),
\end{align}
and the hard function is  given by
\begin{align}\label{eq:H0}
    H^{(0)}(z_A P_A, z_B P_B\to p_C)=\overline{|M^{(0)}|}(z_A P_A, z_B P_B\to p_C)=e_q^2(1-\epsilon) z_A z_B s.
\end{align}
Plugging them into Eq. (\ref{eq:factorizationQ}) gives the same result as Eq. (\ref{eq:sigm0}), hence confirming the validity of factorization at the Born level.

\subsection{Factorization at one loop}
At $O(\alpha_s)$, the off-diagonal amplitude squared  for the general formula in Eq. (\ref{eq:sigmaQCD}) contains both virtual and real diagrams. Instead of evaluating them exactly, we constrict ourselves to showing that expanding them to leading order in $\lambda$ yields the one-loop factorized result given by Eq. (\ref{eq:factorizationQ}).

The virtual diagrams include the quark/antiquark self-energies and the one-loop quark-photon vertex function. In dimensional regularization, the quark/antiquark self-energies vanish due to the cancellation between ultraviolet (UV) and infrared (IR) poles for massless onshell particles. Therefore, one only needs to include the one-loop quark-photon vertex function, which contains both UV and IR divergences. Since its counterterm cancels with that for quark/antiquark self-energies, the UV pole is, effectively, converted into an IR pole and the singularities in the virtual diagrams are of IR origin. Using the Laudau equation~\cite{Landau:1959fi}, one can see that there are three potentially IR divergent regions in the phase space of the virtual gluon: 1) $n_A$-collinear region; 2) $n_B$-collinear region; and 3) the soft region. These regions produce a double IR pole in $\epsilon$. The one-loop quark-photon vertex function is well known, which only depends on $2p_A\cdot p_B$ (see, e.g., Ref. \cite{Sterman:1994ce}). Its explicit form is not relevant for our discussion here, and we write its contribution to the total cross section in the following compact form
\begin{align}\label{eq:sigm1v}
    \frac{d\sigma^{(1)}_r}{d^2\mathbf{b}} =&\frac{\pi}{s}\int\frac{d^2\mathbf{q}}{(2\pi)^2}e^{i\mathbf{q}\cdot\mathbf{b}} \overline{|M^{(1)}_v|^2}(p_A, p_B\to p_C \leftarrow\bar{p}_A, \bar{p}_B)\notag\\
        &\times \delta((p_A+p_B)^2-Q^2)
\end{align}
with $\overline{|M^{(1)}_v|^2}$ the spin-averaged amplitude squared which contains virtual-gluon contributions.

The real amplitude includes two diagrams:
\begin{align}\label{eq:Mr}
    iM^{(1)}_r(p_A, p_B\to p_C, k)\equiv\begin{array}{c} \includegraphics[width=0.5\textwidth]{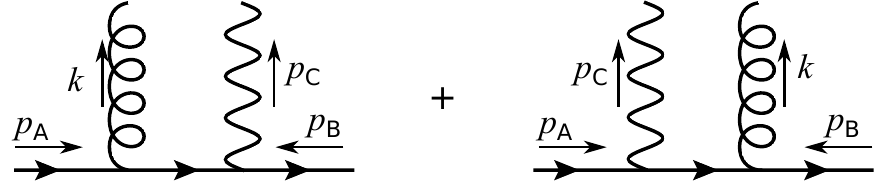}
    \end{array}.
\end{align}
Accordingly, the real correction in perturbative QCD takes the form
\begin{align}\label{eq:sigm1r}
    \frac{d\sigma^{(1)}_r}{d^2\mathbf{b}} =&\frac{\pi}{s}\int\frac{d^2\mathbf{q}}{(2\pi)^2}e^{i\mathbf{q}\cdot\mathbf{b}}\int d\Gamma_k \overline{|M^{(1)}_r|^2}(p_A, p_B\to p_C, k \leftarrow\bar{p}_A, \bar{p}_B)\notag\\
        &\times \delta((p_A+p_B-k)^2-Q^2).
\end{align}
It also contains collinear and soft divergences.

We use the method of regions (see Ref. \cite{Becher:2014oda} for an introduction) to expand the real (Eq. (\ref{eq:sigm1v}))  and virtual (Eq. (\ref{eq:sigm1r})) integrals in each relevant region of the gluon momentum $k^\mu$ in order to verify the factorization formula at one loop. The relevant regions include:

\noindent\underline{1. The hard region: $k^\mu\sim Q$}\\
In this region, upon expanding the off-diagonal amplitude squared, at leading order in $\lambda$ it equals the conventional amplitude squared with $p_i$ and $\bar{p}_i$ both replaced by their design values $P_i$.
As a result, the expansion of the virtual and real integrals in this region gives
\begin{align}\label{eq:hardregion}
    \frac{d\sigma^{(1)}_{h}}{d^2\mathbf{b}}
     =&\delta^{(2)}(\mathbf{b})\sigma^{(1)}(P_A,P_B\to p_C)
\end{align}
with $\sigma^{(1)}$ the conventional one-loop cross section for $q\bar{q}\to\gamma^*$, which can be found in, e.g., Ref. \cite{Sterman:1994ce}. The double poles in virtual and real contributions cancel out but the collinear divergences do not cancel, which produces the $1/\epsilon$ pole in the final result of $\sigma^{(1)}$. If one inserts into the factorization formula the zeroth-order transverse spatial PDFs in Eq. (\ref{eq:T0qq}) and the spin-averaged virtual and real amplitude squared with incoming momenta equal to $P_i$ as the one-loop hard function, one evidently reproduces the same result as Eq. (\ref{eq:hardregion}).

\noindent\underline{2. Two collinear regions: $k^\mu\sim Q(\lambda^2, 1, \lambda)_{n_i\bar{n}_i}$}\\
Expanded in these regions, the virtual diagrams become scaleless integrals and, hence, vanish. As a result, one only needs to consider the real diagrams as shown in Eq. (\ref{eq:Mr}). 

Let us take for example the $n_A$-collinear region in which
\begin{align}\label{eq:scalingA}
    p_A^\mu\sim \bar{p}_A^\mu\sim k^\mu\sim  (\lambda^2, 1, \lambda)_{n_A\bar{n}_A},\qquad
    p_B^\mu\sim \bar{p}_B^\mu\sim  (1, \lambda^2, \lambda)_{n_A\bar{n}_A}.
\end{align}
With the above scaling in mind, expanding the $\delta$ function in Eq. (\ref{eq:sigm1r}) gives
\begin{align}
\delta(z_A s-Q^2) 
\end{align}
with $(1-z_A)\equiv\bar{n}_A \cdot k/\bar{n}_A\cdot P_A$. Then, expand the real amplitude $M_r^{(1)}$ in $\lambda$ as well. After some algebra, we have
\begin{align}
    \overline{|M_{r,n_A}|^2}(p_A, p_B\to p_C, k \leftarrow\bar{p}_A, \bar{p}_B)=e_q^2 (1-\epsilon) s\frac{g^2_s C_F}{2}\frac{N_{qq}(z) \left[4 |\mathbf{k}|^2-(1-z)^2|\mathbf{q}|^2 \right]}{\left|\mathbf{k}+ (1-z)\frac{\mathbf{q}}{2}\right|^2\left|\mathbf{k}- (1-z)\frac{\mathbf{q}}{2}\right|^2}
\end{align}
with
\begin{align}\label{eq:Nqq}
    N_{qq}(z)=\left(1+z^2\right) - \epsilon (1-z)^2.
\end{align}
In this way, one can get the leading-order contribution from the $n_A$-collinear region. In order to verify the factorization formula, we need to show that
\begin{align}\label{eq:sigma1rnA}
    \frac{d\sigma^{(1)}_{r,n_A}}{d^2\mathbf{b}}     &=\frac{\pi}{s}\int\frac{d^2\mathbf{q}}{(2\pi)^2}e^{i\mathbf{q}\cdot\mathbf{b}}\int d\Gamma_k \overline{|M_{r,n_A}|^2}(p_A, p_B\to p_C, k \leftarrow\bar{p}_A, \bar{p}_B)\delta(z_A s-Q^2)\notag\\
        &=\frac{\pi}{s}\int \frac{dz_A}{z_A} \mathcal{T}_{q/q}^{(1)}(\mathbf{b},z_A) H^{(0)}(z_A P_A, P_B\to p_C)\delta(z_A s-Q^2)
\end{align}
with the zeroth-order hard function $H^{(0)}$ given in Eq. (\ref{eq:H0}) and $\mathcal{T}_{q/q}^{(1)}$ the one-loop transverse spatial quark distribution function to be calculated below. The physical meaning of this equation is quite obvious: the quark, recoiling against a radiated gluon, approaches to the antiquark located at $\mathbf{b}$ and then they annihilate into a photon of virtuality $Q^2$.

Let us confirm Eq. (\ref{eq:sigma1rnA}) using the factorization formula in Eq. (\ref{eq:factorizationQ}). Here, we need to calculate the one-loop transverse spatial quark distribution function in the incoming quark. According to its definition in Eq. (\ref{eq:Wq}), one has, in $\bar{n}\cdot A=0$ lightcone gauge,
\begin{align}\label{eq:T1qq}
    \mathcal{T}^{(1)}_{q/q}(\mathbf{r}, z)=\int\frac{d^2\mathbf{q}}{(2\pi)^2}e^{i\mathbf{q}\cdot\mathbf{r}}\int\frac{d^{2-2\epsilon}\mathbf{k}}{(2\pi)^{2-2\epsilon}}\frac{\mathcal{M}}{4\pi (1-z)\bar{n}\cdot P}
\end{align}
with
\begin{align}\label{eq:Ma}
\mathcal{M} &=\begin{array}{c}
     \includegraphics[width=0.4\textwidth]{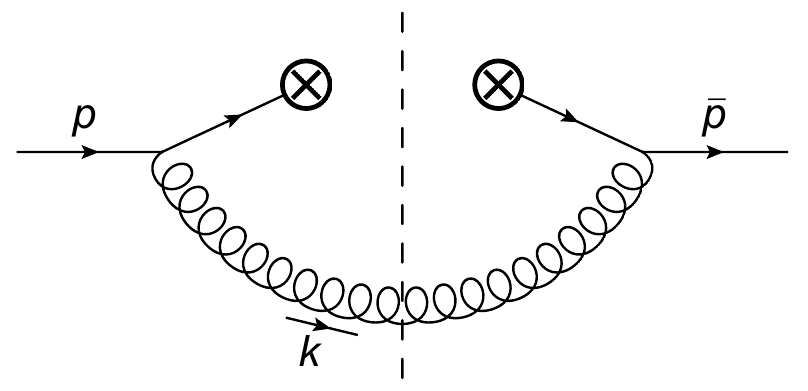}
\end{array} \notag\\
&= g_{s}^{2}C_F \frac{1}{2}\frac{\text{Tr}\left[u^s_{p}\bar{u}^s_{\bar{p}}\slashed{\epsilon}^*_\lambda(k)\left(\slashed{\bar{p}}-\slashed{k}\right)\frac{\slashed{\bar{n}}}{2}\left(\slashed{p}-\slashed{k}\right)\slashed{\epsilon}_\lambda(k)\right]}{\left(p-k\right)^2\left(\bar{p}-k\right)^2}.
\end{align}
Here, the gluon polarization sum in the lightcone gauge is given by
\begin{align}
    \sum\limits_\lambda \epsilon^{\mu}_{\lambda}(k){\epsilon^*}^{\nu}_{\lambda}(k)=-g^{\mu\nu}+\frac{\bar{n}^\mu k^\nu+\bar{n}^\nu k^\mu}{\bar{n}\cdot k}.
\end{align}
The incoming quark momenta in the amplitude ($p\equiv P+q_T/2$) and the conjugate amplitude ($\bar{p}\equiv P-q_T/2$) are both onshell:
\begin{align}
    P^\mu\pm\frac{q_T^\mu}{2}\equiv- \frac{q_T^2}{4 \bar{n}\cdot P} \frac{\bar{n}^\mu}{2}+\bar{n}\cdot P\frac{n^\mu}{2}\pm \frac{q_T^\mu}{2}.
\end{align}
The gluon momentum $k^\mu$ can be decomposed as
\begin{align}
    k^\mu = -\frac{k_T^2}{(1-z)\bar{n}\cdot P}\frac{\bar{n}^\mu}{2}+(1-z)\bar{n}\cdot P \frac{n^\mu}{2}+k_T^\mu.
\end{align}
Accordingly, the denominators in Eq. (\ref{eq:Ma}) are given by
\begin{align}
    \left(P\pm \frac{q_T}{2}-k\right)^2=\frac{\left(k_T\mp (1-z)\frac{q_T}{2}\right)^2}{{1-z}}\sim \lambda^2.
\end{align}

One only needs to keep terms of $O(\lambda^{-2})$ in $\mathcal{M}$. After expanding it according to the scaling in Eq. (\ref{eq:scalingA}), one can easily obtain
\begin{align}\label{eq:MaLO}
   \mathcal{M} = g_{s}^{2}C_F \bar{n}\cdot P\frac{N_{qq}(z) \left[4 |\mathbf{k}|^2-(1-z)^2|\mathbf{q}|^2 \right]}{2\left|\mathbf{k}+ (1-z)\frac{\mathbf{q}}{2}\right|^2\left|\mathbf{k}- (1-z)\frac{\mathbf{q}}{2}\right|^2}
\end{align}
with $N_{qq}(z)$ given in Eq. (\ref{eq:Nqq}). The equality of Eq. (\ref{eq:sigma1rnA}) is confirmed by plugging $\mathcal{M}$ in the above equation (via Eq. (\ref{eq:T1qq})) and the zeroth-order hard function in Eq. (\ref{eq:H0}) into the last line of Eq. (\ref{eq:sigma1rnA}).

The one-loop transverse spatial quark distribution in a fast-moving quark can be calculated analytically. One combines the denominators on the right-hand side of Eq. (\ref{eq:MaLO}) by introducing a Feynman parameter $x$:
\begin{align}
    x\left|\mathbf{k}+ (1-z)\frac{\mathbf{q}}{2}\right|^2+(1-x)\left|\mathbf{k}- (1-z)\frac{\mathbf{q}}{2}\right|^2=\left|\mathbf{\tilde{k}}\right|^2+\Delta
\end{align}
with
\begin{align}
    \mathbf{\tilde{k}}= \mathbf{k}-\frac{1}{2} (1-2 x) (1-z)\mathbf{q},\qquad \Delta = x (1-x) (1-z)^2 |\mathbf{q}|^2.
\end{align}
Then, by changing variables to $\mathbf{\tilde{k}}$, one can easily integrate out $\mathbf{k}$ in Eq. (\ref{eq:T1qq}) and obtain
\begin{align}\label{eq:Wgg1}
        \mathcal{T}^{(1)}_{q/q}(\mathbf{r}, z)&=\frac{\alpha_s C_F}{2\pi}\int_0^1dx\int\frac{d^2\mathbf{q}}{(2\pi)^2}e^{i\mathbf{q}\cdot\mathbf{r}}\frac{ N_{qq}(z)(1-2\epsilon)\Gamma (\epsilon ) \left(\frac{e^{\gamma_E } \mu ^2}{ x(1-x) |\mathbf{q}|^2}\right)^{\epsilon }}{(1-z)^{1+2 \epsilon}}\notag\\
        &=\frac{\alpha_s C_F}{2\pi}\frac{\Gamma(\epsilon)\Gamma^2(1-\epsilon)}{\Gamma(1-2\epsilon)}\frac{N_{qq}(z)}{(1-z)^{1+2 \epsilon}}\int\frac{d^2\mathbf{q}}{(2\pi)^2}e^{i\mathbf{q}\cdot\mathbf{r}}\left(\frac{e^{\gamma_E } \mu ^2}{ |\mathbf{q}|^2}\right)^{\epsilon }\notag\\
        &=\frac{\alpha_s C_F}{2\pi^{\frac{5}{2}}r^2}\cos(\pi\epsilon)\Gamma^2(1-\epsilon)\Gamma\left(\epsilon+\frac{1}{2}\right)\frac{N_{qq}(z)}{(1-z)^{1+2 \epsilon}}\left(e^{\gamma_E } \mu ^2 r^2\right)^{\epsilon },
\end{align}
where we have used the integral in Eq. (\ref{eq:Ixy}) and replaced the bared coupling $g_s^2$ by
\begin{align}
    \frac{g_s^2}{4\pi} = \alpha_s\left(\frac{\mu^2 e^{\gamma_E}}{4\pi }\right)^\epsilon 
\end{align}
with $\mu$ the $\overline{\text{MS}}$ renormalization scale and $\gamma_E$ the Euler constant. At the end, by using
\begin{align}
\frac{1}{(1-z)^{1+2 \epsilon}} = -\frac{1}{2\epsilon}\delta(1-z) + \frac{1}{(1-z)_+},
\end{align}
we have
\begin{align}\label{eq:Tqq1}
     \mathcal{T}^{(1)}_{q/q}(\mathbf{r}, z)=\frac{\alpha_s C_F}{2\pi^2 r^2}\left[ -\left(\frac{1}{\epsilon}+L_T\right)\delta(1-z)  + \frac{1+z^2}{(1-z)_+}\right]
\end{align}
with
\begin{align}
    L_T\equiv2\log\left(\frac{r\mu}{2e^{-\gamma_E}}\right).
\end{align}
The singularity in $\epsilon$ arises from the fact that the virtual-gluon radiation can only contribute to the prefactor in front of $\delta(\mathbf{r})$ and there is no real-virtual cancellation at a nonzero transverse distance $\mathbf{r}$.

Similarly, one can also identify the contribution from the transverse spatial antiquark distribution function of the incoming $\bar{q}$ by expanding the integrand of Eq. (\ref{eq:sigm1r}) in the $n_B$-collinear region. One hence confirms the contributions from one-loop transverse spatial PDFs in the factorization formula.

\noindent\underline{3. The soft region: $k^\mu\sim\lambda Q$}\\
As a consistency check, we show that the contribution from the soft region vanishes. We expand both the diagram for the one-loop spatial quark distribution function in Eq. (\ref{eq:Ma}) and the virtual and real integrals for the general formula given respectively by Eqs. (\ref{eq:sigm1v}) and (\ref{eq:sigm1r}) in this region. The former corresponds to the zero-bin contribution to the spatial PDFs while the latter corresponds to the one-loop correction to the soft function. In both cases, one ends up with scaleless integrals, which vanish in dimensional regularization. Therefore, the soft region is indeed irrelevant.

In summary, we have verified the validity of the factorization formula at one loop: the correction from the one-loop hard function is given by the expansion of the virtual and real integrals for the general formula in the hard region, as given in Eq. (\ref{eq:hardregion}); the correction from one-loop transverse spatial PDFs is equivalent to expanding these virtual and real integrals respectively in the two collinear regions, as given in Eq. (\ref{eq:sigma1rnA}) for the quark collinear region; and the correction from the soft function vanishes.

\section*{Acknowledgements}
The author would like to thank F. Gelis, Y. V. Kovchegov, J. G. Milhano and U. A. Wiedemann for reading through the manuscript and informative comments. The author is also grateful to Y.-T. Chien, R. Rahn, S. S. van Velzen, D. Y. Shao and W. J. Waalewijn for illuminating discussions on aspects of SCET. This work has received support from Xunta de Galicia (Centro singular de investigaci\'on de Galicia accreditation 2019-2022 and Talent Attraction Program), from the European Union ERDF and from the Spanish Research State Agency by “María de Maeztu” Units of Excellence program MDM-2016-0692. The author is also indebted to CERN TH department for continued support when this work was finalized.

\appendix

\section{An integral}
\begin{figure}
\includegraphics[width=0.5\textwidth]{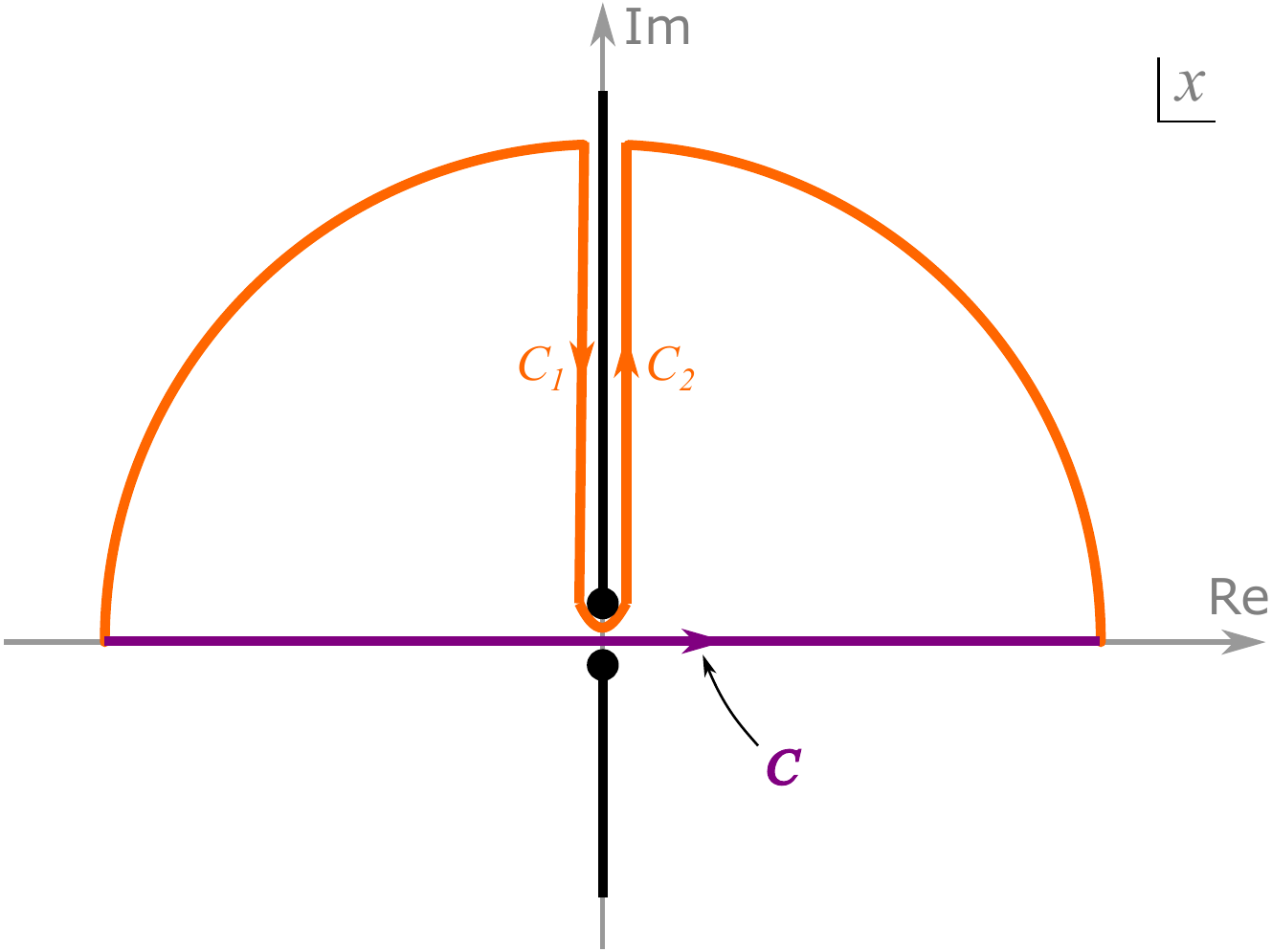}
\caption{Integration contour for $I_x$. The two branch cuts extend respectively from $i0^+$ to $i\infty$ and from $i0^-$ to $-i\infty$.
\label{fig:contour}}
\end{figure}

In this appendix, we evaluate the integral
\begin{align}
    I=\int\frac{d^2 \mathbf{q}}{(2\pi)^2}\frac{e^{i \mathbf{q}\cdot\mathbf{r}}}{(|\mathbf{q}|^2)^{\epsilon}}.
\end{align}
One can choose $\mathbf{r}$ to align with the positive $x$-axis and split the integral into two pieces:
\begin{align}
    I=\frac{1}{(2\pi)^2}\frac{1}{r^{2-2\epsilon}}I_x I_y.
\end{align}
Here,
\begin{align}
    I_x &\equiv \int dx (x^2)^{\frac{1}{2}-\epsilon} e^{i x}\qquad\text{with $x=q_x r$}\notag\\
    &=\int\limits_C dx [(x-i0^+)(x-i0^-)]^{\frac{1}{2}-\epsilon} e^{i x}
\end{align}
with the contour $C$ running from $-\infty$ to $+\infty$. This integral is well-defined only for the range $\frac{1}{2}<\text{Re}~\epsilon < 1$, where one can deform the contour from $C$ to $C_1$+$C_2$ as shown in Fig. \ref{fig:contour}. On $C_1$ the arguments of $x-i0^+$ and $x-i0^-$ are $-i\frac{3\pi}{2}$ and $i\frac{\pi}{2}$ respectively while on $C_2$ they are both equal to $i\frac{\pi}{2}$. Using this fact, one has
\begin{align}
    I_x=i\int_0^\infty dx x^{1-2\epsilon} e^{-x}[e^{i(\frac{1}{2}-\epsilon)\pi}-e^{-i(\frac{1}{2}-\epsilon)\pi}]=-2 \cos(\epsilon\pi)\Gamma(2-2\epsilon).
\end{align}
$I_y$ is defined as
\begin{align}
    I_y\equiv\int dy \frac{1}{(1+y^2)^\epsilon}=\frac{\sqrt{\pi}\Gamma(\epsilon-\frac{1}{2})}{\Gamma(\epsilon)}.
\end{align}
With $I_x$ and $I_y$ evaluated above, one finally has
\begin{align}\label{eq:Ixy}
    I=\int\frac{d^2 \mathbf{q}}{(2\pi)^2}\frac{e^{i \mathbf{q}\cdot\mathbf{r}}}{(|\mathbf{q}|^2)^{\epsilon}}= -\frac{1}{2\pi^{\frac{3}{2}}}\frac{1}{r^{2-2\epsilon}}\cos(\epsilon\pi)\frac{\Gamma(2-2\epsilon)\Gamma(\epsilon-\frac{1}{2})}{\Gamma(\epsilon)}.
\end{align}

\providecommand{\href}[2]{#2}\begingroup\raggedright\endgroup
\end{document}